\documentstyle[preprint,aps,epsfig]{revtex}
\tighten

\newcommand{\Slash}[1]{\!\not\hskip-0.3mm#1}   

\newcommand{\ppp}[1]{%
        \setbox0=\hbox{#1}%
        \kern-.02em\copy0\kern-\wd0
        \kern+.04em\copy0\kern-\wd0
        \kern-.02em\raise.0217em\box0}
\newcommand{\vek}[1]{
        \mathchoice{\mbox{\boldmath$#1$}}%
        {\mbox{\boldmath$#1$}}%
        {\ppp{$\scriptstyle#1$}}%
        {\ppp{$\scriptscriptstyle#1$}}}

\newcommand{\lsim}{$\raisebox{-0.8ex} {$\stackrel{\textstyle <}{\sim}$}$}

\begin{document}  
\preprint{
\vbox{
\hbox{TUM-T39-97-15}
}}

\setcounter{footnote}{1}
\title{\Large\bf Deuteron Spin Structure Functions 
\linebreak
at Small Bjorken-$x$ 
\footnote{Work supported in part by BMBF}}

\bigskip

\author{J.~Edelmann, G.~Piller and  W.~Weise}

\maketitle

\begin{center}
\vspace*{1cm}

Physik Department, Technische Universit\"{a}t M\"{u}nchen, \\
D-85747 Garching, Germany 
\end{center}

\vspace*{3cm}

\begin{abstract} 
We investigate polarized deuteron structure functions at 
small values of the Bjorken variable, $x<0.1$.
In this region contributions from the coherent interaction of 
diffractively excited hadronic states with both nucleons 
become important. 
A proper treatment of this process requires an extension of 
the Glauber-Gribov multiple scattering theory to include 
spin degrees of freedom. 
In the kinematic domain of current fixed target 
experiments we observe that shadowing effects in  
$g_1^d$ are approximately twice as large as for 
the unpolarized structure function $F_2^d$.
Furthermore at $x<0.1$ the tensor structure function $b_1$ 
is found to receive significant contributions from coherent 
double scattering.


\end{abstract}

\setcounter{footnote}{0}

\newpage

\section{Introduction}

In recent years polarized deep-inelastic scattering experiments have 
become 
a major topic at all high energy lepton beam facilities.
They aim primarily at the investigation of the spin structure of hadrons. 
Here the use of proton targets has led to detailed information  
about the spin structure function $g_1^p$. 
In addition the corresponding  
neutron structure function $g_1^n$ has  been explored. 
Combined with proton data the latter is crucial for the extraction 
of the flavor singlet combination of polarized quark distributions, 
and for testing the fundamental Bjorken sum rule. 

To investigate neutron structure functions nuclear targets are needed.
In current experiments
at CERN \cite{SMC}, SLAC \cite{E144,E143} and HERMES \cite{HERMES_g1}, 
$^{3}$He and deuteron targets are used. 
For an accurate extraction of neutron structure 
functions a detailed knowledge of nuclear effects is 
therefore required. 
At moderate and large values of the Bjorken variable,
$x>0.2$, such effects can be traced back to nuclear 
binding and Fermi motion (see e.g. \cite{binding1} 
and references therein). 
At small $x<0.1$ corrections due to 
coherent multiple scattering processes become important.
In unpolarized deep-inelastic scattering 
these are responsible for nuclear shadowing which has been established 
in recent  experiments at FNAL \cite{E665d,E665} and CERN \cite{NMC}
as a leading twist effect.

In this paper we study polarized deep-inelastic scattering 
from deuterium at small $x$. 
Important contributions to the corresponding cross section come from 
mechanisms in which the exchanged virtual photon scatters diffractively 
from one of the nucleons in the target and produces hadronic intermediate
states which subsequently interact with the second nucleon.
Coherent double scattering is described 
by the Glauber-Gribov multiple scattering theory  
which we extend in the present work 
to include spin degrees of freedom. 

We estimate shadowing effects 
in the deuteron spin structure function $g_1^d$. 
The corrections thus obtained turn out to be approximately 
a factor two  larger than 
those for the unpolarized structure function $F_2^d$. 
Nevertheless, as  compared to other uncertainties in the
extraction of $g_1^n$ from deuteron data, these  
corrections are relatively small.

Because of its spin-$1$ nature 
the deuteron is characterized by 
additional structure functions as compared to a free nucleon.
Amongst those the yet unmeasured structure function 
$b_1$ \cite{HoJaMa89,Mil_b1,SaSc90}
will be investigated at HERMES \cite{HERMES_b1}. 
Model calculations suggest that $b_1$ is very small at 
moderate and large $x>0.2$ (see e.g. \cite{b1_large_x}). 
At small $x<0.1$ dominant contributions 
to $b_1$ result from coherent double scattering processes.
Here the magnitude of $b_1$ reaches around $2\%$ 
of the unpolarized nucleon structure function $F_1^N$. 
This interesting effect is entirely due 
to an interference between the $S$- and 
$D$-state component in the deuteron wave function.

First investigations of double scattering contributions 
to polarized deuteron structure functions have been  
performed in \cite{KhHo93} within a Regge model. 
However, in that work an unrealistic deuteron wave function has been 
used neglecting the $D$-state component.   
At $x<0.1$ the latter turns out to be crucial for a proper 
estimate of the structure function $b_1$ 
as found  in \cite{EdPiWe97,NNSc97} 
(see also \cite{GeWi75,Stri95}).
Shadowing effects for spin structure functions have  been 
discussed recently within the framework of a simple model,  
especially for $^3$He targets \cite{FrGuSt96}. 
The magnitude of the observed effect agrees well with the 
results presented here.

This paper is organized as follows: in Sec.II we 
recall the relationship between structure 
functions and photon helicity amplitudes. 
An extension of the Glauber-Gribov multiple scattering series 
to spin degrees of freedom is presented in Sec.III, where we also 
derive the single and double scattering 
contributions to deuteron structure functions. 
In Sec.IV we estimate shadowing effects in the spin structure function 
$g_1^d$ and discuss  implications for the extraction 
of the neutron structure function $g_1^n$.
Results for $b_1$ are presented in Sec.V. 
We summarize in Sec.VI.

\section{Structure Functions and Helicity Amplitudes}
In inclusive deep-inelastic lepton scattering one probes the 
hadronic tensor
\begin{eqnarray} \label{eq:hadronic_T}
W_{\mu \nu}(p,q,s)
&=& \frac{1}{4\pi} \int d^4\xi\, e^{iq\cdot \xi}
\left\langle p,s \left| J_{\mu}(\xi)  J_{\nu}(0)
                 \right|p,s \right\rangle
\end{eqnarray}
of the target.
Here $J^{\mu}$ is the electromagnetic current.  
The four-momenta of the target and the exchanged photon are labeled by 
$p^{\mu}$ and $q^{\mu}=(q_0,\vek q)$, respectively. 
The spin vector $s^{\mu}$ is orthogonal to the target
momentum, $p\!\cdot\!s=0$, and normalized such that $s^2=-M_T^2$,  
where $M_T$ is the invariant mass of the  target.
For  a spin-1/2 target $s^{\mu}$ directly represents the target 
polarization,
while for a spin-1 target it is defined in terms of the 
polarization vectors ${\cal E}_H$ as 
$s_{\alpha}(H)
= -i \varepsilon_{\alpha\beta\gamma\delta}
     \,{\cal E}_{H}^{\beta *}\,{\cal E}_{H}^{\gamma} p^{\delta}$,
where $H=0,+1,-1$ (abbreviated as $0,+,-$) denotes the spin projection 
along
the quantization axis.

According to the optical theorem inclusive deep-inelastic lepton 
scattering can be described in terms of the forward scattering of a 
virtual 
photon. 
The corresponding Compton amplitude is given by: 
\begin{eqnarray} \label{eq:Compton_T}
T_{\mu \nu}(p,q,s)
&=&  i \int d^4\xi\, e^{iq\cdot \xi}
\left\langle p,s \left| T\left( J_{\mu}(\xi) J_{\nu}(0)
                        \right)
                 \right|p,s \right\rangle.
\end{eqnarray}
A comparison with Eq.(\ref{eq:hadronic_T}) 
gives:
\begin{equation} \label{eq:opt_theorem}
W_{\mu\nu} = \frac{1}{2\pi} Im\,T_{\mu\nu}.
\end{equation}
Using the Compton tensor one can  define 
photon-target forward helicity amplitudes (Fig.1):  
\begin{equation} \label{eq:hel_amp}
{\cal A}_{jH,j'H'} = e^2\varepsilon_{j'}^{*\mu} T_{\mu\nu}(H',H)  
\varepsilon_{j}^{\nu},
\end{equation} 
with the electromagnetic coupling $e^2/4\pi  = 1/137$. 
The helicities of the incoming and scattered photon
are labeled by $j$ and $j'$. 
Choosing the photon momentum in the longitudinal direction, 
$q^{\mu} = (q_0,{\bf 0_{\perp}}, q_z)$ with $q_z > 0$,   
gives for the  photon  polarization 
vectors $\varepsilon_+^{\mu} = (0,-1,-i,0)/\sqrt{2},\,
\varepsilon_-^{\mu} = (0,1,-i,0)/\sqrt{2}$ and 
$\varepsilon_0^{\mu} = (q_z,0,0,q_0)/\sqrt{-q^2}$. 
Furthermore $H$ and $H'$ specify  
the target helicities before and after the interaction.

Lorentz covariance, parity and time reversal 
invariance, hermiticity and current conservation imply that 
the hadronic tensor of the free nucleon is expressed 
in terms of four structure functions $F_1^N, F_2^N$ and $g_1^N, g_2^N$
which depend on the momentum transfer $Q^2 = - q^2$ and 
$x_T = Q^2/2 \nu$ with $\nu = p\cdot q$ 
\footnote{We use the notation $x_T$  for the Bjorken 
variable of a target nucleus, while $x$ denotes that of the 
nucleon as usual, i.e. $x= \frac{Q^2}{2 M q_0}$ in the lab frame.}:
\begin{eqnarray}\label{hadten}
W^N_{\mu\nu}&=&
-g_{\mu\nu}\,F_1^N +
\frac{p_{\mu}p_{\nu}}{\nu}\,F_2^N 
+\frac{i}{\nu}\varepsilon_{\mu\nu\lambda\sigma}
q^{\lambda}s^{\sigma}\,g_1^N+
\frac{i}{\nu^2}\varepsilon_{\mu\nu\lambda\sigma}q^{\lambda}(p\cdot q
s^{\sigma}-s\cdot q p^{\sigma})\,g_2^N.
\end{eqnarray}
With the nucleon helicities $H, H'= +, - = 
\uparrow, \downarrow$ we obtain from 
Eqs.(\ref{eq:opt_theorem},\ref{eq:hel_amp},\ref{hadten}) for the 
spin-averaged
structure functions:
\begin{mathletters}
\label{eq:F^N_all}
\begin{eqnarray}
F_1^N&=&\frac{1}{4\pi e^2}
\left(Im \,{\cal A}^{\gamma^*N}_{+\downarrow,+\downarrow}+
Im \,{\cal A}^{\gamma^*N}_{+\uparrow,+\uparrow}
\right),\label{f1helamp}\\
F_2^N&=&
\frac{x}{2 \pi e^2 \kappa_N}
\left(Im \,{\cal A}^{\gamma^*N}_{+\downarrow,+\downarrow}+
Im \,{\cal A}^{\gamma^*N}_{+\uparrow,+\uparrow}
+ 2 Im \,{\cal A}^{\gamma^*N}_{0\uparrow,0\uparrow}\right),
\end{eqnarray}
and for the spin-dependent ones:
\begin{eqnarray}
g_1^N &=&\frac{1}{4\pi e^2 \kappa_N}
\left(Im \,{\cal A}^{\gamma^*N}_{+ \downarrow,+ \downarrow}-
Im \,{\cal A}^{\gamma^*N}_{+\uparrow,+ \uparrow}+\sqrt{2(\kappa_N-1)}
Im \,{\cal A}^{\gamma^*N}_{+\downarrow,0\uparrow}\right),
\\
g_2^N &=&\frac{1}{4\pi e^2\kappa_N}
\left(Im \,{\cal A}^{\gamma^*N}_{+ \uparrow,+ \uparrow}-
Im \,{\cal A}^{\gamma^*N}_{+\downarrow,+\downarrow}
+ \frac{2}{\sqrt{2(\kappa_N-1)}}
Im \,{\cal A}^{\gamma^*N}_{+\downarrow,0\uparrow}
\right).
\label{g1helamp}
\end{eqnarray} 
\end{mathletters}
Here $\kappa_N=1+{M^2Q^2}/{\nu^2}$ 
with the nucleon mass $M$.

For deuterium, as for any spin-1 target, the hadronic tensor 
has eight structure functions. 
Four of them are proportional to  Lorentz structures 
as found for free nucleons (\ref{hadten}). Of
the remaining structure functions  $b_1, b_2$,$\Delta$ and $b_3$, the
first three   
occur at leading twist while $b_3$ is suppressed at large $Q^2$  
\cite{HoJaMa89,SaSc90}:
\begin{eqnarray} \label{hadten_d}
&&W^d_{\mu\nu}=
-g_{\mu\nu}\,F_1^d +
\frac{p_{\mu}p_{\nu}}{\nu}\,F_2^d
+\frac{i}{\nu}\varepsilon_{\mu\nu\lambda\sigma}
q^{\lambda}s^{\sigma}\,g_1^d+
\frac{i}{\nu^2}\varepsilon_{\mu\nu\lambda\sigma}q^{\lambda}(p\cdot q\,
s^{\sigma}-s\cdot q p^{\sigma})\,g_2^d  
\nonumber\\
&&- \left(g_{\mu\nu} b_1 - \frac{p_{\mu}p_{\nu}}{\nu} b_2\right) 
\,\left(\frac{M_d^2}{\kappa_d \nu^2} q\cdot {\cal E}  
\,q\cdot {\cal E}^* - \frac{1}{3}\right) 
\nonumber \\
&&+ 
\frac{\Delta}{2}
\left\{
\left(-g_{\mu\nu} + \frac{2x_T}{\kappa_d\nu} p_{\mu}p_{\nu} \right)  
\left(\frac{M_d^2}{\kappa_d \nu^2} q\cdot {\cal E}  \,q\cdot {\cal E}^* - 
1\right)
+
\left[
\left({\cal E}_{\mu} - \frac{q\cdot {\cal E}}{\kappa_d\nu}p_{\mu}\right)
\left({\cal E}^*_{\nu} - 
\frac{q\cdot {\cal E}^*}{\kappa_d\nu}p_{\nu}\right)
+ (\mu\leftrightarrow \nu)
\right]
\right\}
\nonumber\\
&&+b_3 \frac{\kappa_d-1}{\sqrt{\kappa_d}\nu} 
\left[p_{\mu} \,q\cdot {\cal E}^* \left({\cal E}_{\nu} - 
\frac{q\cdot {\cal E}}{\kappa_d \nu} p_{\nu} \right) 
+ p_{\mu} \,q\cdot {\cal E} 
\left({\cal E}^*_{\nu} - 
\frac{q\cdot {\cal E}^*}{\kappa_d\nu}p_{\nu}\right)
+ (\mu\leftrightarrow \nu)
\right].
\nonumber \\
&&
\end{eqnarray}
Here we use $\kappa_d=1+{M_d^2Q^2}/{\nu^2}$ and omit the spin indices 
$H$ in the deuteron polarization vector ${\cal E}$.
The relations between helicity amplitudes and structure functions 
are  obtained again from 
Eqs.(\ref{eq:opt_theorem},\ref{eq:hel_amp},\ref{hadten_d}): 
\begin{mathletters}
\label{eq:F^d_all}
\begin{eqnarray}
F_1^d &=& \frac{1}{6\pi e^2} 
\left(
Im\,{\cal A}^{\gamma^* d}_{++,++} +  Im\,{\cal A}^{\gamma^*d}_{+-,+-}  
+ Im\,{\cal A}^{\gamma^*d}_{+0,+0} 
\right),  
\\
F_2^d &=& \frac{x_T}{3\pi e^2\kappa_d} 
\left(
Im\,{\cal A}^{\gamma^*d}_{++,++} +  Im\,{\cal A}^{\gamma^*d}_{+-,+-} + 
Im\,{\cal A}^{\gamma^*d}_{+0,+0} 
+ 2 Im\,{\cal A}^{\gamma^*d}_{0+,0+} + Im\,{\cal A}^{\gamma^*d}_{00,00} 
\right),  
\\
g_1^d &=& \frac{1}{4\pi e^2\kappa_d} 
\left(
Im\,{\cal A}^{\gamma^*d}_{+-,+-} -  Im\,{\cal A}^{\gamma^*d}_{++,++}  
+ \sqrt{\kappa_d-1}(Im\,{\cal A}^{\gamma^*d}_{+0,0+} + 
Im\,{\cal A}^{\gamma^*d}_{+-,00})
\right),  
\\
g_2^d &=& \frac{1}{4\pi e^2\kappa_d} 
\left(
Im\,{\cal A}^{\gamma^*d}_{++,++} -  Im\,{\cal A}^{\gamma^*d}_{+-,+-}  
+ \frac{1}{\sqrt{\kappa_d-1}} 
(Im\,{\cal A}^{\gamma^*d}_{+0,0+} +Im\,{\cal A}^{\gamma^*d}_{+-,00})
\right),
\\  
b_1 &=& -\frac{1}{4\pi e^2} 
\left(
Im\,{\cal A}^{\gamma^*d}_{++,++} +  Im\,{\cal A}^{\gamma^*d}_{+-,+-}  - 
2 Im\,{\cal A}^{\gamma^*d}_{+0,+0} 
\right),  
\\
b_2 &=& -\frac{x_T}{2 \pi e^2\kappa_d} 
\left(
Im\,{\cal A}^{\gamma^*d}_{++,++} +  Im\,{\cal A}^{\gamma^*d}_{+-,+-}  - 
2 Im\,{\cal A}^{\gamma^*d}_{+0,+0} 
+ 2Im\,{\cal A}^{\gamma^*d}_{0+,0+} -2Im\,{\cal A}^{\gamma^*d}_{00,00}  
\right),  
\\
\Delta &=& \frac{1}{2 \pi e^2}
Im\,{\cal A}^{\gamma^*d}_{+-,-+}, 
\\
b_3 &=& \frac{1}{4\pi e^2\sqrt{\kappa_d(\kappa_d-1)}}
(Im\,{\cal A}^{\gamma^*d}_{+0,0+} - Im\,{\cal A}^{\gamma^*d}_{+-,00}).
\end{eqnarray}
\end{mathletters}%
The scaling property  of the structure 
functions (\ref{eq:F^d_all}) requires that the amplitudes  
${\cal A}^{\gamma^*d}_{+0,0+}$ and 
${\cal A}^{\gamma^*d}_{+-,00}$ drop at least as $1/\sqrt{Q^2}$.

\section{Deep-inelastic scattering from deuterons at small 
$\lowercase{x}$}
\label{Sec:Deu_small_x}

{}From the previous discussion we find that in 
the scaling limit, which we use throughout, the structure functions  
$F^d_1, F^d_2, g^d_1$ and $b_1$  are determined 
by the helicity conserving amplitudes  
${\cal A}^{\gamma^*d}_{+H} \equiv {\cal A}^{\gamma^*d}_{+H,+H}$. 
These can be split into single and double scattering parts as 
sketched  in Fig.2:
\begin{equation}
{\cal A}^{\gamma^* d}_{+H}={\cal A}^{(1)}_{+H} + {\cal A}^{(2)}_{+H}. 
\end{equation}
In the single scattering term the photon interacts incoherently with 
the proton or neutron of the target. 
The double scattering term involves interactions in which 
both nucleons take part.
In the following we derive the corresponding 
amplitudes extending the Glauber-Gribov multiple scattering theory 
\cite{Glauber59,Gribov69,DS}
to include spin degrees of freedom.

\subsection{Single scattering}

We  first focus on single scattering contributions  
to the structure functions $F_{1(2)}^d, g_1^d$ and $b_1$.
The corresponding Compton amplitude reads (Fig.2a): 
\begin{eqnarray}\label{eq:A_1} 
i{\cal A}^{(1)}_{+H}&=&-\int\frac{d^4 p}{(2\pi)^4}
{\cal E}^{\alpha*}_{H}\varepsilon^{\mu*}_+
Tr\left[\Gamma_{\beta}(p_d,p)\frac{i}{\Slash p_d-\Slash p
-M+i\varepsilon} \bar{\Gamma}_{\alpha}(p_d,p)
\right.\nonumber\\
&&\left. \hspace*{2cm}
\times\frac{i}{\Slash p
-M+i\varepsilon}\;i\hat{t}^{\gamma^*p}_{\mu\nu}(p,q)\frac{i}{\Slash
p -M+i\varepsilon}\right]
{\cal E}^{\beta}_{H}\varepsilon^{\nu}_+ +[p\leftrightarrow n],
\end{eqnarray}
We describe the scattering process in the laboratory 
frame where the deuteron with four-momentum $p_d^{\mu} = (M_d,\bf 0)$ 
is at rest. 
The integration in (\ref{eq:A_1}) runs over the  momentum 
$p^{\mu}=(p^0,\vek p)$ of the interacting nucleon.
The polarization vectors of the transverse photon and 
the deuteron  are denoted by $\varepsilon_+$ and ${\cal E}_H$, 
respectively.
All information about deuteron structure is absorbed in the 
vertex functions $\Gamma$, such that the free proton and neutron 
propagators remain.  
The interaction of the photon with the bound nucleon 
is described by the reduced amplitude  $\hat{t}^{\gamma^*N}$ 
which is related to the physical photon-nucleon forward scattering 
amplitudes   in (\ref{eq:F^N_all}) as follows:
\begin{equation}\label{redhl}
{\cal A}^{\gamma^*N}_{+h}=\varepsilon^{\mu *}_+\;\bar{u}(p,h)\;
\hat{t}^{\gamma^* N}_{\mu\nu}(p,q)\;u(p,h)\;\varepsilon^{\nu}_+.
\end{equation}
Here $u(p,h)$ is the Dirac spinor of a nucleon with momentum $p$ and 
helicity $h$. 

One can perform the energy integration in (\ref{eq:A_1}) assuming 
that all relevant poles are included in the nucleon propagators. 
We neglect modifications of the individual 
nucleon amplitudes ${\cal A}^{\gamma^* N}$ 
due to nuclear binding and Fermi motion. 
They  are relevant only at moderate and large $x$ 
(see e.g. \cite{binding1,binding2}),
whereas our primary interest in this paper is the region $x<0.1$.
Next we perform a non-relativistic expansion of the nucleon propagators 
in (\ref{eq:A_1}), keeping  only the leading terms in $|\vek p|/M$. 
In the non-relativistic limit we identify the vertex functions 
$\Gamma$  with non-relativistic deuteron wave functions $\psi_H$. 
We find (for details see Appendix A):
\begin{eqnarray} \label{eq:A_1wf}
{\cal A}^{(1)}_{+H}&=&\int d^3r \,
\psi^{\dag}_{H}(\vek{r}) 
\left( P^p_{\uparrow}{\cal A}^{\gamma^*p}_{+\uparrow}+
P^p_{\downarrow}{\cal A}^{\gamma^*p}_{+\downarrow} \right)
\,\psi_{H}(\vek{r}) +[p\leftrightarrow n]. 
\end{eqnarray}
The  operators $P^N_{\uparrow(\downarrow)}$ act 
on the deuteron wave function and project  onto a proton  
or neutron with helicity  $+$ or $-$.
The coordinate-space wave function of the deuteron is:
\begin{eqnarray} \label{psi}
\psi_{H}(\vek{r})&=&\frac{1}{\sqrt{4\pi}}\left[\frac{u(r)}{r}+
\frac{v(r)}{r}\frac{1}{\sqrt{8}}\hat{S}_{12}(\hat{\vek{r}})
\right]\chi_{H},
\end{eqnarray}
where $\hat{S}_{12}(\hat{\vek{r}}) = 
3 (\vek \sigma_p\cdot \vek r) (\vek \sigma_n\cdot \vek r)/r^2 - 
\vek \sigma_p\cdot \vek \sigma_n$ is  the tensor operator  
with  $r=|\vek r|$, and $\chi_{H}$ denotes the  
spin wave function of the triplet proton-neutron pair.
The deuteron $S$- and $D$-state components are determined by the 
radial wave functions $u$ and 
$v$, respectively. 
They are normalized according to 
$\int_0^{\infty} dr \,[u^2(r)+v^2(r)] =1$, while the $D$-state 
probability is given by $\omega_D = \int_0^{\infty} dr \,v^2(r)$.

Inserting the deuteron wave function in  Eq.(\ref{eq:A_1wf}) 
and working out the projection operators 
for the different deuteron polarizations, one finds:
\begin{mathletters}
\begin{eqnarray}
{\cal A}^{(1)}_{++}&=&\left(1-\frac{3}{4}\omega_D\right)
\left({\cal A}^{\gamma^*p}_{+\uparrow}+
{\cal A}^{\gamma^*n}_{+\uparrow}\right)
+\frac{3}{4}\omega_D\left({\cal A}^{\gamma^*p}_{+\downarrow}
+{\cal A}^{\gamma^*n}_{+\downarrow}\right),\label{a+}\\
{\cal A}^{(1)}_{+-}&=&\left(1-\frac{3}{4}\omega_D\right)
\left({\cal A}^{\gamma^*p}_{+\downarrow}+
{\cal A}^{\gamma^*n}_{+\downarrow}
\right)
+\frac{3}{4}\omega_D\left({\cal A}^{\gamma^*p}_{+\uparrow}
+{\cal A}^{\gamma^*n}_{+\uparrow}\right),\label{a-1}\\
{\cal A}^{(1)}_{+0}&=&\frac{1}{2}
\left({\cal A}^{\gamma^*p}_{+\uparrow}+{\cal A}^{\gamma^*p}_{+\downarrow}
+{\cal A}^{\gamma^*n}_{+\uparrow}+
{\cal A}^{\gamma^*n}_{+\downarrow}\right).\label{a0}
\end{eqnarray}
\end{mathletters}
With Eqs.(\ref{eq:F^N_all},\ref{eq:F^d_all}) 
the single scattering contributions to the deuteron 
structure functions become: 
\begin{mathletters}
\begin{eqnarray}
F_1^d&=&F_1^p+F_1^n = 2 F_1^N,\\
g_1^d&=&\left(1-\frac{3}{2}\omega_D\right)(g_1^p+g_1^n) 
=(2 - 3 \,\omega_D)\,g_1^N,\\
b_1^&=&0.
\end{eqnarray}
\end{mathletters}
We end up with 
the incoherent sum of the corresponding proton and neutron structure 
functions.
The polarized structure function $g_1^d$ includes 
the depolarization factor $(1-3\omega_D/2)$. 
It accounts for the fact that the deuteron and the interacting nucleon 
can be polarized 
in opposite directions if the deuteron is in a $D-$state.
Note that $b_1$ vanishes for single scattering since there is no such
structure function for the individual spin-$1/2$ particles.

\subsection{Double scattering}
\label{sec:DS}

At small values of the Bjorken variable, $x<0.1$, double scattering 
contributes
significantly to photon-deuteron Compton scattering 
and consequently to deuteron structure functions.
In this process the virtual photon diffractively produces a hadronic
intermediate state on the first nucleon, which subsequently interacts 
with the
second nucleon.
In the case of unpolarized scattering 
destructive interference  of the single and double scattering amplitudes 
leads to the observed nuclear shadowing 
(see e.g. \cite{PiRaWe95}). 

The helicity dependent double scattering amplitude reads (see Fig.2b):
\begin{eqnarray} 
\label{eq:A2_ds}
i {\cal A}^{(2)}_{+H}&=&
-\sum_X\int\frac{d^4 p}{(2\pi)^4}
\int\frac{d^4k}{(2\pi)^4}
{\cal E}^{\alpha*}_{H}\varepsilon^{\mu*}_+
\frac{-i(g^{\rho\sigma}-p_X^{\rho}p_X^{\sigma}/M_X^2)}
{p_X^2-M_X^2+i\varepsilon}
\nonumber\\
&\times& Tr\left[\Gamma_{\beta}
(p_d,p)\,
\frac{i}{\Slash p_d-\Slash p -M+i\varepsilon}
i\,\hat{t}^{Xn\to \gamma^*n}_{\mu\rho}\,
\frac{i}{\Slash p_d-\Slash p- \Slash k-M+i\varepsilon}
\bar{\Gamma}_{\alpha}(p_d,p)\,
\right.\nonumber\\
&&\hspace*{0.5cm}
\times \left.\frac{i}{\Slash p +\Slash k-M+i\varepsilon}i\,
\hat{t}^{\gamma^*p\to Xp}_{\sigma\nu}\,
\frac{i}{\Slash p-M+i\varepsilon}\,
\right]
{\cal E}^{\beta}_{H}\varepsilon^{\nu}_+\nonumber\\
&+&[p\leftrightarrow n].
\end{eqnarray}
Here $p$ denotes the four-momentum of the first interacting nucleon  
and $k$ the  momentum transfer. 
The sum is taken over all diffractively excited hadronic states $X$ 
which carry photon quantum numbers, 
an invariant mass  $M_X$ and  a four-momentum  $p_X =q-k$. 
Their contributions to double scattering are   
determined by the reduced amplitudes 
$\hat t^{\gamma^* N\rightarrow XN}$ 
which are related to the  
diffractive (virtual) photoproduction amplitudes   
of hadronic states $X$ from nucleons by:
\begin{equation}\label{freie2}
T^{NX}_{+h}(k)=\varepsilon^{\sigma *}_{X+} \,
\bar{u}(p+k,h)\hat{t}^{\gamma^*N\to XN}_{\sigma\nu}
u(p,h) \,\varepsilon^{\nu}_{+},
\end{equation}
where $\varepsilon_{X+}$ denotes the transverse polarization 
vector of the produced hadronic intermediate states. 
We choose the photon momentum in the longitudinal direction,  
$q^{\mu}=(q_0,{\bf 0}_{\perp}, \sqrt{q_0^2 + Q^2})$. 
As in  single scattering we perform the energy 
integration in (\ref{eq:A2_ds})  
assuming that all relevant poles are in the propagators. 
We then take into account only the leading non-relativistic contribution 
to ${\cal A}^{(2)}$. 
Furthermore 
we assume that the diffractive amplitudes $\hat t$ depend on 
the transverse momentum transfer only. 
Following the steps given in Appendix B this leads to: 
\begin{eqnarray} \label{eq:A_2_ds}
{\cal A}^{(2)}_{+H}&=&\frac{i}{4Mq_0}\sum_X
\int\frac{d^2k_{\perp}}{(2\pi)^2}\int d^2 b\,
e^{i\vek{k}_{\perp}\cdot\vek{b}}
\int_{-\infty}^0dz\,
e^{i\frac{z}{\lambda}}\nonumber\\
&\times&\psi^{\dag}_{H}(\vek{r})
\left(
P^p_{\uparrow}T^{pX}_{+\uparrow}(k)+P^p_{\downarrow}
T^{pX}_{+\downarrow}(k)\right)\otimes\left(
P^n_{\uparrow}T^{nX}_{+\uparrow}(k)+P^n_{\downarrow}
T^{nX}_{+\downarrow}(k)\right)
\psi_{H}(\vek{r}) \nonumber\\
&+&[p\leftrightarrow n].
\end{eqnarray}
In analogy with hadron-hadron high-energy collisions  
\cite{DonLan89} the real part of  diffractive production amplitudes is 
supposed to be small. We therefore use in the following 
$T^{NX}\approx i Im\, T^{NX}$ and consider only terms which 
contribute to the imaginary part of the 
double scattering amplitude ${\cal A}^{(2)}$.

For the double scattering 
contribution to be significant,  the 
longitudinal propagation length
\begin{equation} \label{cohlength}
\lambda\simeq\frac{2 q_0}{M_X^2+Q^2}=\frac{1}{x\;M}
\left(\frac{Q^2}{M^2_X + Q^2}\right)
\end{equation}
of a diffractively excited hadron must exceed the size of the 
deuteron target, 
$d=\langle r^2\rangle_d^{1/2}\approx 4\,fm$. 
At large $Q^2$ the important intermediate states are those 
with $M_X^2 \sim Q^2$ 
(see e.g. Refs.\cite{Gribov69,PiRaWe95,FraStr89}). 
Hence the condition $\lambda > d$ is fulfilled for $x<0.03$ 
in agreement with the observed shadowing  effect in 
unpolarized deep-inelastic scattering \cite{E665d,E665,NMC}. 

After projecting onto proton and neutron states with definite 
helicity one finds:

\begin{mathletters}\label{eq:A_dH^(2)_int}
\begin{eqnarray} \label{eq:App_etc}
{\cal A}^{(2)}_{++}& \approx &\frac{i}{4Mq_0}\sum_X
\int\frac{d^2k_{\perp}}{(2\pi)^2}\, 
\,S_+(\vek k_{\perp},1/\lambda) \,
\,T^{pX}_{+\uparrow}(k)
  T^{nX}_{+\uparrow}(k),
\\
{\cal A}^{(2)}_{+-}& \approx &\frac{i}{4Mq_0}\sum_X
\int\frac{d^2k_{\perp}}{(2\pi)^2}\, 
\,S_-(\vek k_{\perp},1/\lambda) \,
  T^{pX}_{+\downarrow}(k)
  T^{nX}_{+\downarrow}(k),
\\
{\cal A}^{(2)}_{+0}& \approx &\frac{i}{8Mq_0}\sum_X
\int\frac{d^2k_{\perp}}{(2\pi)^2}\, 
\,S_0(\vek k_{\perp},1/\lambda)\, 
\left[ T^{pX}_{+\uparrow}(k)
       T^{nX}_{+\downarrow}(k) + 
       T^{pX}_{+\downarrow}(k)
       T^{nX}_{+\uparrow}(k) \right],
\end{eqnarray}
\end{mathletters}
with the helicity dependent deuteron form factor
\begin{equation} \label{eq:FF_H}
S_{H}(\vek k)=\int
d^3r\,|\psi_{H}(\vek r )|^2 e^{i\vek k \cdot \vek r}.
\end{equation}
In Eq.(\ref{eq:A_dH^(2)_int}) we have neglected contributions which 
vanish for diffractive production processes 
in the forward direction. 
At small scattering angles $\theta$ the omitted terms 
are suppressed by typical factors $(\sin \theta/\cos \theta)^2$. 
Note in addition that the neglected contributions are 
proportional to the square of the deuteron 
$D$-state wave function and 
therefore numerically insignificant in any case.  
The complete  expressions for the helicity amplitudes 
${\cal A}^{(2)}$ can be found in Appendix B.

For the following discussion we approximate 
the dependence of the 
diffractive production amplitudes on the 
momentum transfer $t = k^2 \approx - \vek k_{\perp}^2$ by: 
\begin{equation} \label{eq:T_ampl_approx}
T^{NX}(k) \approx e^{- B \,\vek k_{\perp}^2 /2 }  
\,T^{NX}. 
\end{equation}
with the forward amplitude $T^{NX} \equiv T^{NX}(\vek k = 0)$.  
For possible small momentum transfers an exponential suppression 
of $T^{NX}$ with rising $|t|$  
is certainly justified  and supported by experiment 
\cite{Bauer,Chapin85,Breitweg97,Crit97}. 
Furthermore we  finally investigate our results for deuteron spin 
structure functions in the kinematic range of 
fixed target experiments at CERN (NMC, COMPASS), 
FNAL (E665) and  DESY (HERMES). 
Here in average  moderate momentum transfers, $Q^2\,\lsim \,3\,GeV^2$, 
are accessible at $x<0.1$. 
Various data on diffractive leptoproduction in 
this kinematic region \cite{Bauer,Chapin85,Breitweg97,Crit97} suggest  
an average slope $B \simeq (6 \dots 10) \,GeV^{-2}$.
Note that the limit $B=0$ corresponds to the approximation that the 
diffractive production of intermediate hadronic states 
proceeds in the forward direction, 
i.e. at a fixed impact parameter $\vek b =0$ (\ref{eq:A_2_ds}).

The double scattering amplitudes ${\cal A}^{(2)}$ can  now 
be expressed in terms of the integrated form factors: 
\begin{equation} \label{eq:F_H}
{\cal F}_{H}(1/\lambda,B)=
\int\frac{d^2k_{\perp}}{(2\pi)^2}\, 
\,S_H(\vek k_{\perp},1/\lambda) \, e^{-B \,\vek k_{\perp}^2}. 
\end{equation}
In Fig.3 we present  ${\cal F}_{H}$ for realistic wave functions 
as obtained from the Paris \cite{LaLoRi80} and Bonn \cite{Bonn}  
nucleon-nucleon  potentials.    
We observe a significant dependence on the 
deuteron polarization. 
Furthermore we find ${\cal F}_H \approx constant$ for 
$\lambda > \langle r^2\rangle^{1/2}_d \approx 4\,fm$. 
To investigate the influence of the diffractive slope $B$ 
we compare results for $B=7\,GeV^{-2}$ and $B=0$. 
We find that the form factors for unpolarized  
deuterium,
\begin{equation} \label{eq:F_unpol}
{\cal F} = \frac{1}{3} \left({\cal F}_+ + {\cal F}_- + {\cal F}_0\right),
\end{equation}
and for transverse polarization,  ${\cal F}_+$,  
are not very sensitive to the exact value of the diffractive 
slope.
In the  examples shown in Fig.3 they vary by maximally $15\%$. 
The situation is different for tensor polarization. 
For $\lambda > 4\,fm$ the corresponding form factor  
\begin{equation} \label{eq:F_02}
{\cal F}_{T} = \frac{1}{2} 
\left({\cal F}_+ + {\cal F}_- - 2 {\cal F}_0\right)
\end{equation}
increases approximately by a factor of three when $B=0$ is chosen 
instead of  $B=7\,GeV^{-2}$. 
The reason is that the deuteron tensor 
form factor  $S_{T} = S_+ + S_- - 2 S_0$  
receives significant contributions from rather large 
momenta $|\vek k| \simeq (200$--$400)\,MeV$. 
In this kinematic region $S_{T}$ varies 
slowly with $t \sim - \vek k_{\perp}^2$. 
As a consequence 
the forward approximation, $B=0$, is poorly 
justified in this case.\footnote{We 
thank N.N. Nikolaev and M.I. Strikman for discussions 
related to this issue.}
In addition note that the differences between form factors 
derived from the Bonn and the Paris potential become 
small if a realistic slope is used.

Combining Eqs.(\ref{eq:A_dH^(2)_int},\ref{eq:T_ampl_approx},\ref{eq:F_H})
finally gives:
\begin{mathletters}\label{eq:A_dH^(2)_final}
\begin{eqnarray} \label{eq:A2_H+-0}
{\cal A}^{(2)}_{++}&=&\frac{i}{4 Mq_0}\sum_X \,
T^{pX}_{+\uparrow} \; T^{nX}_{+\uparrow}
{\cal F}_{+}(1/\lambda,B),\label{A2+}
\\
{\cal A}^{(2)}_{+-}&=&\frac{i}{4Mq_0}\sum_X \,
T^{pX}_{+\downarrow} \; T^{nX}_{+\downarrow}\,
{\cal F}_{-}(1/\lambda,B),\label{A2-}
\\
{\cal A}^{(2)}_{+0}&=&\frac{i}{8Mq_0}\sum_X 
\left(T^{pX}_{+\downarrow} \; T^{nX}_{+\uparrow}
+ T^{pX}_{+\uparrow}\; 
T^{nX}_{+\downarrow}\right){\cal F}_{0}(1/\lambda,B).\label{A20}
\end{eqnarray}
\end{mathletters}
In combination with Eq.(\ref{eq:F^d_all}) we then obtain the 
double scattering corrections to all deuteron structure functions.

For the unpolarized structure function 
$F_1^d = F_1^p + F_1^n + \delta F_1$ we have:
\begin{eqnarray} \label{eq:deltaF_1^d}
\delta F_1&=&
\frac{1}{6\pi e^2}
\left[Im {\cal A}_{++}^{(2)}+Im {\cal A}_{+-}^{(2)} + 
Im {\cal A}^{(2)}_{+0} \right],\\
&=&
\frac{1}{32 \pi e^2 M q_0}
\sum_X\left\{{\cal F}(1/\lambda,B)
\left(T^{pX}_{+\uparrow}T^{pX}_{+\uparrow}+
T^{pX}_{+\downarrow}T^{pX}_{+\downarrow}
+T^{nX}_{+\uparrow}T^{nX}_{+\uparrow}+
T^{nX}_{+\downarrow}T^{nX}_{+\downarrow}
\right)\right.
\label{delf1}
\\
&&\hspace*{2.6cm}-{\cal F} (1/\lambda,B)\left[(T^{pX}_{+\uparrow}-
T^{nX}_{+\uparrow})^2
+(T^{pX}_{+\downarrow}-T^{nX}_{+\downarrow})^2\right]
\nonumber\\
&&\hspace*{2.6cm}-\left.\frac{2}{3}{\cal F}_{0}(1/\lambda,B)
(T^{pX}_{+\uparrow}-T^{pX}_{+\downarrow})
(T^{nX}_{+\uparrow}-T^{nX}_{+\downarrow})\right\}.\label{eq:deltaF_1^d2}
\nonumber
\end{eqnarray}
If we ignore  the isospin and spin dependent combinations 
proportional to
\begin{eqnarray}\label{Spin-Isospin}
(T^{pX}_{+\uparrow}&-&T^{nX}_{+\uparrow})^2, 
(T^{pX}_{+\downarrow}-T^{nX}_{+\downarrow})^2
\,\,\mbox{and}\,\, 
(T^{pX}_{+\uparrow}-T^{pX}_{+\downarrow}) 
(T^{nX}_{+\uparrow}-T^{nX}_{+\downarrow}), 
\end{eqnarray}
we arrive at the well known result \cite{Gribov69}
\begin{equation} \label{eq:F_1^d_Gribov}
\delta F_1=-\frac{2 Q^2}{e^2 x}
\int_{4 m_{\pi}^2}^{W^2} dM_X^2
\left(
\left.\frac{d^2\sigma^{\gamma^*_T N}_{\downarrow}}
{dM_X^2 dt}\right|_{t=0}
+\left.\frac{d^2\sigma^{\gamma^*_TN}_{\uparrow}}
{dM_X^2 dt}\right|_{t=0}\right)
{\cal F}(1/{\lambda},B),
\end{equation}
written in terms of 
the forward diffractive hadron production cross section in the 
collision  of transverse virtual photons with  nucleons 
at $t=(p_X-q)^2 \approx 0$.  
The individual helicity dependent diffractive cross sections are:  
\begin{equation}
16\pi \int_{4m_{\pi}^2}^{W^2} dM_X^2 
\left.\frac{d^2\sigma_{h}^{\gamma^*_T N}}{dM_X^2 dt}
\right|_{t\approx 0} 
= 
\frac {1}{8 M^2 q_0^2}\sum _X 
\left(|T^{pX}_{+h}|^2 + |T^{nX}_{+h}|^2 \right). 
\end{equation}
In Sec.\ref{sec:IVB} we investigate all terms of Eq.(\ref{delf1}) 
as they 
contribute to $\delta F_1$ within a simple model. 
The corrections to the conventional result, 
Eq.(\ref{eq:F_1^d_Gribov}), turn out to be small indeed.
In Fig.4 we present the shadowing correction for the unpolarized 
structure function $F_2^d/2F_2^N$ as measured by the E665 
collaboration \cite{E665d}. 
At $x\ll 0.1$ one finds $\delta F_2/2 F_2^N  \approx 
\delta F_1/2F_1^N\approx - 0.03$, 
with  large experimental errors.
Note however that the quoted value is consistent with 
an analysis of the $A$-dependence of nuclear shadowing 
in Ref.\cite{shad_A}. 
For later estimates of double scattering effects in deuteron 
spin structure functions we use the fit to the data shown in 
Fig.4. 
Corresponding theoretical calculations can be found for example in  
Ref.\cite{shad_deu}. 

Following Eqs.(\ref{eq:F^d_all},\ref{eq:A_dH^(2)_final}), 
the double scattering to the polarized structure function
$g_1^d=(1-{3}\,\omega_D/2 )(g_1^p+g_1^n)+\delta g_1 $ is: 
\begin{eqnarray} \label{eq:dg1_1}
\delta g_1&=&\frac{1}{4\pi e^2}
\left[Im {\cal A}_{+-}^{(2)}-
Im {\cal A}_{++}^{(2)}\right] 
\\
&=& \label{eq:deltag_1^d}
\frac{1}{32\pi e^2 M q_0 }\sum_X
{\cal F}_{+}(1/\lambda,B)
\left[T^{pX}_{+\downarrow}\,T^{pX}_{+\downarrow}
-T^{pX}_{+\uparrow}\,T^{pX}_{+\uparrow}
+T^{nX}_{+\downarrow}\,T^{nX}_{+\downarrow}-
T^{nX}_{+\uparrow}\,T^{nX}_{+\uparrow}\right.
\nonumber\\
&& \hspace*{4cm}+\left.(T^{pX}_{+\uparrow}-
T^{nX}_{+\uparrow})^2-
(T^{pX}_{+\downarrow}-T^{nX}_{+\downarrow})^2\right].
\end{eqnarray}
If one assumes that  spin averaged diffractive amplitudes 
are equal for protons and neutrons, i.e.
$T^{pX}_{+\uparrow}+T^{pX}_{+\downarrow}=
T^{nX}_{+\uparrow}+T^{nX}_{+\downarrow}$, 
the third and fourth terms in Eq.(\ref{eq:deltag_1^d}) vanish. 
In Regge phenomenology this is 
guaranteed by the isospin independent pomeron exchange which 
dominates diffraction at small $x$ (see e.g. \cite{Goulia83}). 
We then end up with: 
\begin{eqnarray} \label{eq:shadg1_diff}
\delta g_1 = -\frac{2 Q^2}{e^2 x}
\int_{4 m_{\pi}^2}^{W^2} dM_X^2
\left(
\left.\frac{d^2\sigma^{\gamma^*_T N}_{\downarrow}}
{dM_X^2 dt}\right|_{t=0}
-\left.\frac{d^2\sigma^{\gamma^*_TN}_{\uparrow}}
{dM_X^2 dt}\right|_{t=0}\right)
{\cal F}_{+}(1/\lambda,B).
\end{eqnarray}
Now the  difference of the helicity dependent 
diffractive production cross sections enters, 
whereas the unpolarized case, 
Eq.(\ref{eq:F_1^d_Gribov}), involves  
the sum of both.

Finally we study  
the structure function $b_1$ which does not receive 
contributions from single scattering 
as long  as nuclear binding and Fermi motion 
are neglected.
Here double scattering gives:
\begin{eqnarray}
b_1&=& -\frac{1}{4\pi e^2} 
\left( Im\,{\cal A}^{(2)}_{++} +  Im\,{\cal A}^{(2)}_{+-}  
- 2 Im\,{\cal A}^{(2)}_{+0} 
\right),  \nonumber \\
&=&
\frac{1}{32 \pi e^2 M q_0}
\sum_X \left\{{\cal F}_{0}(1/\lambda,B)
\left[T^{pX}_{+\uparrow}\,T^{pX}_{+\uparrow}+
T^{pX}_{+\downarrow}\,T^{pX}_{+\downarrow}
+T^{nX}_{+\uparrow}\,T^{nX}_{+\uparrow}
+T^{nX}_{+\downarrow}\,T^{nX}_{+\downarrow}
\right.\right.
\\
&&\hspace*{4.5cm} \left.-(T^{pX}_{+\uparrow}-
T^{nX}_{+\uparrow})^2- 
(T^{pX}_{+\downarrow}-T^{nX}_{+\downarrow})^2
\right.\nonumber 
\\
&&\left.
\hspace*{4.5cm}
-2(T^{pX}_{+\uparrow}-T^{pX}_{+\downarrow})
(T^{nX}_{+\uparrow}-T^{nX}_{+\downarrow})\right]
\nonumber\\
&&\hspace*{2.5cm}-{\cal F}_{+}(1/\lambda,B)
\left[T^{pX}_{+\uparrow}\,T^{pX}_{+\uparrow}
+T^{pX}_{+\downarrow}\,T^{pX}_{+\downarrow}+
T^{nX}_{+\uparrow}\,T^{nX}_{+\uparrow}
+T^{nX}_{+\downarrow}\,T^{nX}_{+\downarrow}\right.\nonumber\\
&&\hspace*{4.5cm}\left.\left.-(T^{pX}_{+\uparrow}-
T^{nX}_{+\uparrow})^2-
(T^{pX}_{+\downarrow}-T^{nX}_{+\downarrow})^2\right]\right\}.
\nonumber
\end{eqnarray}
With the approximations (\ref{Spin-Isospin}) we obtain:
\begin{equation} \label{eq:b1_ds}
b_1=\frac{4 Q^2}{e^2 x}
\int_{4 m_{\pi}^2}^{W^2} dM_X^2
\left.\frac{d^2\sigma^{\gamma^*_TN}}
{dM_X^2 dt}\right|_{t=0}
{\cal F}_{T}(1/\lambda,B),
\end{equation}
and observe that $b_1$ 
is proportional to the difference of the 
deuteron form factors (\ref{eq:F_02}) for transverse and longitudinal 
target polarizations.
This contribution is entirely 
driven by the interference of the 
deuteron $S$- and $D$-state component  
as can be seen from  Eq.(\ref{eq:AppB_SH}). 
Note that our result in Eq.(\ref{eq:b1_ds}) 
implies a violation of the quark model sum rule \cite{NNSc97}
\begin{equation}
\int_0^1 dx\, b_1(x,Q^2) = 0, 
\end{equation}
which has been discussed in Ref.\cite{b1_SR}.

\section{Shadowing in $\lowercase{g}_1^{\lowercase{d}}$}
\label{sec:IVB} 

Shadowing in the deuteron spin 
structure function $g_1^d$ is proportional to 
the difference of polarized diffractive virtual photoproduction cross 
sections, see Eq.(\ref{eq:shadg1_diff}).
These quantities are  not measured up to now.
Nevertheless it is possible to estimate the shadowing correction 
$\delta g_1$ to an accuracy which is sufficient for the 
extraction of the neutron structure function $g_1^n$ from 
recent experimental data \cite{SMC,E143}. 
In the kinematic region of current experiments it is 
legitimate to neglect nuclear  binding and Fermi motion corrections 
\cite{binding2}. 
To leading order in the shadowing correction $\delta g_1/\delta g_1^N$ 
one then obtains:
\begin{equation} \label{eq:g1n_est}
g_1^n \simeq  
\frac{g_1^d}{1 - \frac{3}{2}\omega_D} 
\left( 1 - \frac{\delta g_1}{2 g_1^N 
\left(1 - \frac{3}{2}\omega_D\right)} 
\right)- 
g_1^p. 
\end{equation}
Recent fixed target experiments find 
$|g_1^d| <  |g_1^p|$ \cite{SMC,E143}. 
The sensitivity of $g_1^n$ to uncertainties 
in the nuclear depolarization or shadowing is therefore suppressed. 
Note that in this respect the use of $^3$He targets is expected 
to be less favorable \cite{FrGuSt96}.

\subsection{Upper limit for shadowing in $g_1^d$}

An  upper limit for shadowing in $g_1^d$ can be obtained by 
directly comparing 
$\delta g_1$ and $\delta F_1$  in terms of the photon-deuteron 
helicity amplitudes (\ref{eq:deltaF_1^d},\ref{eq:dg1_1}). 
Note that at small $x$ 
their imaginary parts have the same sign. 
The reason is that at $x\ll 0.1$ large values 
of the coherence length  are relevant 
(see Eq.(\ref{cohlength})). 
For $\lambda > 4\,fm$ the longitudinal deuteron 
form factors (\ref{eq:F_H}) which enter in  Eq.(\ref{eq:A_dH^(2)_final})
are all positive. 
We can then apply the Schwartz inequality 
\begin{equation}
\left|
Im {\cal A}^{(2)}_{++}+
Im {\cal A}^{(2)}_{+-}+
Im {\cal A}^{(2)}_{+0}
\right|\ge \left|Im {\cal A}^{(2)}_{++}+Im {\cal
A}^{(2)}_{+-}\right|\ge\left|Im {\cal
A}^{(2)}_{+-} -Im {\cal A}^{(2)}_{++}\right|,
\end{equation}
and obtain:
\begin{equation} \label{eq:ub_dg1}
\frac{\left|\delta g_1\right|}{F_1^N}\le \frac{3}{2} 
\frac{\left|\delta F_1\right|}{F_1^N}
\approx \frac{3}{2}\frac{\left|\delta F_2\right|}{F_2^N}.
\end{equation}
Given the E665 data on  $F_2^d/2 F_2^N$ shown in Fig.4 
we conclude from Eqs.(\ref{eq:g1n_est},\ref{eq:ub_dg1}) that 
uncertainties due to the shadowing correction $\delta g_1$ are small.
They are within the experimental errors of recent data analyses 
\cite{SMC,E143}.
However, once accurate data for $g_1^d$ and $g_1^p$ 
become available at $x< 0.01$ 
the upper bound (\ref{eq:ub_dg1}) is not helpful anymore due to 
the strong rise of $F_1^N$ at small $x$ (see e.g. \cite{FNHERA}).


\subsection{Model calculation}
In this section we investigate  double scattering 
contributions to deuteron structure functions 
in the framework of a simple model. 
In a laboratory frame 
description of deep-inelastic scattering at small $x\ll 0.1$, 
the exchanged virtual photon first converts to a  
hadronic state $X$   which then interacts with the target. 
In this kinematic region the photon-nucleon helicity 
amplitudes ${\cal A}^{\gamma^* N}_{+h}$ can 
be described  by hadron-nucleon amplitudes 
averaged over all hadronic states present in the 
photon wave function. 
Here hadronic states with invariant mass $M_X^2 \sim Q^2$ dominate 
as one can find 
from a  spectral analysis of  photon-nucleon amplitudes 
(e.g. \cite{Gribov69,PiRaWe95,FraStr89}). 
We therefore approximate the photon-nucleon helicity  
amplitudes by effective hadron-nucleon amplitudes
$\overline {\cal A}_{+h}^{XN}$ 
which depend weakly on kinematic parameters, times a factor which 
incorporates the leading $Q^2$-dependence:
\begin{equation}  \label{eq:A_av}
{\cal A}_{+h}^{\gamma^* N} \approx 
{e ^2} \frac{C(Q^2,x)}{Q^2} \overline {{\cal A}}_{+h}^{XN}
\end{equation}
In the case of unpolarized scattering an effective hadron-nucleon 
cross section $\overline \sigma^{XN} = \frac{1}{2M q_0} 
Im\,\overline{\cal A}^{XN}\approx 17\,mb$ 
reproduces the measured shadowing in the nuclear 
structure functions $F_{2}^A$ \cite{FrGuSt96}.
Furthermore one finds, that in the kinematic domain of current 
fixed target experiments, the function $C(Q^2,x)$ depends only weakly on 
$Q^2$ and $x$ \cite{PiRaWe95,FraStr89}, while 
the factor $1/Q^2$ ensures scaling of 
deep-inelastic structure functions as can be seen from 
Eqs.(\ref{eq:F^N_Xall},\ref{eq:F^d_all_X}).

Replacing the amplitudes  (\ref{eq:F^N_all}) by the effective 
ones of Eq.(\ref{eq:A_av}) yields for the nucleon structure functions:  
\begin{mathletters}
\label{eq:F^N_Xall}
\begin{eqnarray}
F_1^N&=&\frac{C}{4 \pi Q^2}
\left(Im \,\overline{\cal A}^{XN}_{+\downarrow}+
Im \,\overline{\cal A}^{XN}_{+\uparrow}
\right),\\
g_1^N &=&\frac{C}{4 \pi Q^2}
\left(Im \,\overline{\cal A}^{XN}_{+ \downarrow}-
Im \,\overline{\cal A}^{XN}_{+\uparrow}\right).
\end{eqnarray} 
\end{mathletters}
The deuteron structure functions 
(\ref{eq:F^d_all}) are expressed in a similar way in 
terms of effective hadron-deuteron amplitudes:
\begin{mathletters}
\label{eq:F^d_all_X}
\begin{eqnarray}
F_1^d &=& \frac{C}{6 \pi Q^2} 
\left(
Im\,\overline{\cal A}^{Xd}_{++} +  
Im\,\overline{\cal A}^{Xd}_{+-}  + 
Im\,\overline{\cal A}^{Xd}_{+0} 
\right),  
\\
g_1^d &=& \frac{C}{4 \pi Q^2} 
\left(
Im\,\overline{\cal A}^{Xd}_{+-} +  
Im\,\overline{\cal A}^{Xd}_{++}  
\right),  
\\  
b_1 &=& -\frac{C}{4 \pi Q^2} 
\left(
Im\,\overline{\cal A}^{Xd}_{++} +  
Im\,\overline{\cal A}^{Xd}_{+-}  - 
2 Im\,\overline{\cal A}^{Xd}_{+0} 
\right).  
\end{eqnarray}
\end{mathletters}
The next step is to  express the deuteron  
amplitudes $\overline{\cal A}^{Xd}$ in terms of nucleon amplitudes.
For the double scattering contributions  
the corresponding relations are identical to 
those of Eq.(\ref{eq:A_dH^(2)_final}),  
if one substitutes  the diffractive amplitudes 
$T^{NX}$ by  $\overline{\cal A}^{XN}$. 
Using Eqs.(\ref{eq:F^N_Xall}) we obtain: 
\begin{mathletters}\label{eq:ds_allX}
\begin{eqnarray} 
\label{eq:ds_F1X}
\delta F_1(x,Q^2) &=& \frac{- \pi x}{C(Q^2,x)}
\left[{\cal F}(1/\lambda,B) F_1^N(x,Q^2)^2 
\right.\nonumber \\
&&
\left.
\hspace*{1.5cm} +  \frac{1}{3}\left(
{2}{\cal F}_{+}(1/\lambda,B)- 
{\cal F}_{0}(1/\lambda,B) \right) 
g_1^p(x,Q^2) g_1^n(x,Q^2) 
\right], 
\\
\delta g_1(x,Q^2) &=& \frac{- 2 \pi x}{C(Q^2,x)}
{\cal F}_{+}(1/\lambda,B) F_1^N(x,Q^2) g_1^N(x,Q^2),
\\
\label{eq:ds_b1X}
b_1(x,Q^2) &=& \frac{\pi x}{C(Q^2,x)}
\left[ {\cal F}_{T}(1/\lambda,B)\,F_1^N(x,Q^2)^2  
\right.\nonumber \\
&&
\left.+
\left({\cal F}_{+}(1/\lambda,B) + {\cal F}_0(1/\lambda,B)\right)
g_1^p(x,Q^2) g_1^n(x,Q^2) 
\right], 
\end{eqnarray}
\end{mathletters}
with $\lambda \approx  1/2Mx$. 
In Eqs.(\ref{eq:ds_F1X},\ref{eq:ds_b1X}) 
the spin and isospin combinations (\ref{Spin-Isospin}) turn out 
to be proportional to the product of the proton and neutron spin 
structure 
functions $g_1$. 
In the kinematic range of current fixed target experiments they 
amount to less than $5\%$ of the dominant  contribution to $\delta F_1$
and  $b_1$, the one is proportional to the square of the 
unpolarized nucleon structure function $F_1^N$. 
As a consequence the standard expression for  
shadowing (\ref{eq:F_1^d_Gribov}) 
and Eq.(\ref{eq:b1_ds}) for $b_1$ at small $x$ 
seem to be  good approximations. 

A comparison of shadowing  for  unpolarized and 
polarized structure functions in Eqs.(\ref{eq:ds_allX}) gives:
\begin{eqnarray} \label{eq:g1_F1}
\frac{\delta g_1}{g_1^N} \approx {\cal R}_{g_1} 
\frac{\delta F_1}{F_1^N} 
\approx 
{\cal R}_{g_1} \frac{\delta F_2}{F_2^N}, 
\quad \mbox{with}\quad 
{\cal R}_{g_1} = 2 \frac{{\cal F}_{+}(2Mx,B)}{{\cal F}(2Mx,B)}.
\end{eqnarray}
At small $x$ and  $B = 7\,GeV^{-2}$ we find ${\cal R}_{g_1} = 2.2$ 
for both, the Paris and Bonn nucleon-nucleon potentials 
\cite{LaLoRi80,Bonn}.
(In the forward approximation, i.e. $B=0$, one has  
${\cal R}_{g_1} = 2.7$ for the Paris, 
and ${\cal R}_{g_1} = 2.4$ for the Bonn potential.) 
In Fig.5 we present the ratio $\delta g_1/g_1^N$ using 
the measured $F_2^d/2 F_2^N$ from Fig.4 \cite{E665d}.
One should note that as $x$ decreases the data for the shadowing ratio 
$F_2^d/2F_2^N$  are taken at decreasing 
values of the average momentum transfer 
$\overline Q^2$ \cite{E665d}. 
Therefore our results for $\delta g_1/g_1^N$ shown in Fig.5 correspond, 
strictly speaking,  to the fixed target kinematics of 
E665 \cite{E665d} 
which is not far from the kinematics of SMC \cite{SMC}. 
{}From Eq.(\ref{eq:g1n_est}) we finally find that 
the shadowing correction 
amounts to less than $5\%$ of the experimental error on 
$g_1^n$ for the SMC \cite{SMC}  and E143 \cite{E143} data analysis.

\section{The tensor structure function ${\lowercase{ b}_1}$ 
         at small $\lowercase{x}$}

The shadowing correction $\delta F_{1}$ for the unpolarized 
structure function  and 
the deuteron tensor structure function $b_1$ are directly related. 
At $x\ll 0.1$ 
the propagation lengths (\ref{cohlength}) of 
diffractively excited hadrons which dominate double scattering exceed 
the deuteron size
$\lambda > \langle r^2\rangle_d^{1/2} \approx 4\,fm$.
Here, as shown in Fig.3, the deuteron form factors saturate, 
i.e. ${\cal F}_H (1/\lambda < 0.25\,fm^{-1},B) \approx {\cal F}_H (0,B)$.
A comparison of Eqs.(\ref{eq:F_1^d_Gribov}) and (\ref{eq:b1_ds}) then 
gives:
\begin{equation} \label{eq:b1_est}
b_1= {\cal R}_{b_1} \,\delta F_1, 
\quad \mbox{with}\quad 
{\cal R}_{b_1} = -\frac{{\cal F}_{T}(0,B)}
{{\cal F}(0,B)}.
\end{equation}
{}Using $B=7\,GeV^{-2}$ we obtain from the Paris nucleon-nucleon 
potential \cite{LaLoRi80} ${\cal R}_{b_1} = -0.33$, while the Bonn 
one-boson-exchange potential \cite{Bonn} gives
${\cal R}_{b_1} = -0.29$.  
A variation of the diffractive slope $B$ by $30\%$ 
leads to a change of the ratio ${\cal R}_{b_1}$  by maximally $20\%$.
In Fig.6 we present $b_1$ as obtained from Eq.(\ref{eq:b1_est}), 
using the fit for  $F_2^d/2F_2^N$ from Fig.4, 
together with  the empirical information on 
$F_1^N$ \cite{E665FN,R}. 

For the ratio of structure functions $b_1/F_1^d$, which is given 
by an asymmetry of inclusive polarized deuteron cross 
sections, $\sigma_{+H}^{\gamma^* d} \sim Im\,{\cal A}_{+H}^{\gamma^* d}$,  
we find (\ref{eq:F^d_all}): 
\begin{equation}
\frac{b_1}{F_1^d} 
= - \frac{3}{2}\,
\frac{\sigma^{\gamma^* d}_{++} +  \sigma^{\gamma^* d}_{+-}  - 
2 \sigma^{\gamma^* d}_{+0}}   
{\sigma^{\gamma^* d}_{++}+\sigma^{\gamma^* d}_{+-}+
\sigma^{\gamma^* d}_{+0}}
\approx 
{\cal R}_{b_1} \frac{\delta F_1}{2 F_1^N} \approx 0.01,  
\end{equation}
i.e. $b_1$ amounts to around $1\%$ of the unpolarized deuteron 
structure function $F_1^d$ or, equivalently, to $2\%$ of $F_1^N$. 
This result agrees with an early estimate in Ref.\cite{Stri95}.
Note that the result  shown here corresponds again to the 
kinematics of E665 \cite{E665d}.
It is therefore relevant with regard to possible future experiments at 
HERMES \cite{HERMES_b1} and eventual COMPASS \cite{COMPASS}.

\section{Summary}
We have studied  nuclear effects in the 
polarized deuteron structure functions 
$g_1^d$ and $b_1$ at small values of the Bjorken variable, $x<0.1$, 
where the diffractive photo-excitation of hadronic 
states on a target nucleon and their 
subsequent interaction with the second nucleon becomes important.
In order to 
describe these coherent  double scattering processes 
in polarized deep-inelastic scattering 
we have extended the Glauber-Gribov multiple scattering theory 
by including  spin degrees of freedom. 
We find that shadowing effects in $g_1^d$ are by a factor two larger 
than those for the unpolarized structure function $F_2^d$. 
Nevertheless they are of minor importance for the extraction of the 
neutron structure function $g_1^n$ from current data on $g_1^d$.
In this respect deuterium seems to be a more favorable target than  
$^3$He.  

We observe that $b_1$ at small $x$  receives  
large contributions from coherent double scattering. 
They come from an interference of the $S$- and  
$D$-state components of the deuteron 
wave function and break the simple quark model sum rule 
which suggests a vanishing first moment of $b_1$. 
At $x<0.1$,  the magnitude of $b_1$ reaches around $2\%$  
of the unpolarized structure function $F_1^N$.

\acknowledgements

We would like to thank N.N. Nikolaev, M. Sargsian 
and M.I. Strikman for  helpful comments and discussions.
This work was supported in part by
BMBF.

\newpage

\section{Appendix}
Here  we outline the derivation of the single 
and double scattering  amplitudes (Eqs.(\ref{eq:A_1wf},\ref{eq:A_2_ds})) 
in  
Sec.\ref{Sec:Deu_small_x}. 
For this purpose we  extend the Glauber-Gribov multiple scattering 
theory \cite{Glauber59,Gribov69,DS} to include spin degrees 
of freedom.

\subsection{Single Scattering}
\label{App_A}

We start out from the single scattering Compton amplitude in 
Eq.(\ref{eq:A_1}):
\begin{eqnarray}\label{Eq.51} 
i{\cal A}^{(1)}_{+H}&=&-\int\frac{d^4 p}{(2\pi)^4}
{\cal E}^{\alpha*}_{H}\varepsilon^{\mu*}_+
Tr\left[\Gamma_{\beta}(p_d,p)\frac{i}{\Slash p_d-\Slash p
-M+i\varepsilon} \bar{\Gamma}_{\alpha}(p_d,p)
\right.\nonumber\\
&&\left. \hspace*{2cm}
\times\frac{i}{\Slash p
-M+i\varepsilon}\;i\hat{t}^{\gamma^*p}_{\mu\nu}(p,q)\frac{i}{\Slash
p -M+i\varepsilon}\right]
{\cal E}^{\beta}_{H}\varepsilon^{\nu}_+ +[p\leftrightarrow n],
\end{eqnarray}
where $\Gamma$ refers to the dpn-vertex and $\hat{t}$ is the reduced 
amplitude
defined in Eq.(\ref{redhl}).
In the laboratory frame which we use throughout, the deuteron 
four-momentum is:
\begin{equation}
p_d^{\mu}=
\left(M_d,{\bf 0}\right)=\left(2(M-B),{\bf 0}\right),
\end{equation}
where $B$ denotes the deuteron binding energy per nucleon.
We perform the energy integration in Eq.(\ref{Eq.51}) 
assuming that all relevant poles are
given by the nucleon propagators.
Neglecting anti-nucleon degrees of freedom this leads to the replacement:
\begin{equation} \label{mass shell}
\frac{1}{(p_d-p)^2-M^2+i\varepsilon}
\to-i\pi\frac{\delta(p_d^0-p^0-E_p)}{E_p},
\end{equation}
with $E_p=\sqrt{M^2+\vek{p}^2}$. We then arrive at:
\begin{equation} \label{single amp2}
{\cal A}^{(1)}_{+H}=
\left.\int\frac{d^3 p}{(2\pi)^3 {2E_p}}\,
\frac{{\cal E}^{\alpha*}_{H}\varepsilon^{\mu*}_+ \,Tr [\dots] \,
{\cal E}^{\beta}_{H}\varepsilon^{\nu}_+}
{\left(p^2-M^2+i\varepsilon\right)^2}\,
\right |_{p_d^0-p^0=E_p} +\,[p\leftrightarrow n],
\end{equation}
with 
\begin{equation} \label{eq:Tr[...]}
Tr [\dots] = 
Tr\left[\Gamma_{\beta}\,(\Slash p_d -\Slash
p+M)\bar{\Gamma}_{\alpha}\,
(\Slash p+M)
\,\hat{t}^{\gamma^*p}_{\mu\nu}(p,q)
(\Slash p+M)
\right].
\end{equation}
In the numerator of Eq.(\ref{single amp2}) we neglect
terms of order  $\vek{p}^2/M^2$ and replace 
$\protect{\Slash p}+M  \approx  \sum_{h}u(p,h) \bar{u}(p,h)$ etc. in 
Eq.(\ref{eq:Tr[...]}), 
where $u(p,h)$ is the Dirac spinor of a nucleon with momentum $p$ 
and helicity $h$. 
This gives:
\begin{eqnarray}
Tr [\dots ] \approx 
&&
\sum_{h,h',h''} Tr \left[ \Gamma_{\beta}\,u(p_d-p,h')
\bar{u}(p_d-p,h')\bar{\Gamma}_{\alpha}
\,u(p,h'')
\bar{u}(p,h'')\,\hat{t}^{\gamma^*p}_{\mu\nu}(p,q)u(p,h)
\bar{u}(p,h)\,
\right].
\end{eqnarray}
Next we apply helicity conservation and introduce the photon-nucleon 
amplitude as in Eq.(\ref{redhl}):
\begin{eqnarray}
\varepsilon^{\mu*}_+ \,Tr [\dots] \, \varepsilon^{\nu}_+
\approx 
&& 
\sum_{h,h'} Tr \left[\Gamma_{\beta}u(p_d-p,h') \bar{u}(p_d-p,h')
\bar{\Gamma}_{\alpha}\, 
\,u(p,h) \bar{u}(p,h)\,
{\cal A}^{\gamma^* p}_{+h}
\right].
\end{eqnarray}
We then expand the nucleon Dirac spinors in $|\vek{p}|/M$
and keep the leading terms, e.g. 
\begin{eqnarray}
u(p,h)&=&\sqrt{E_p+M}\left(\array{c}\chi_h\\\frac{\vek{\sigma}\cdot
\vek{p}}{E_p+M}\chi_h\endarray\right) 
\approx  \sqrt{2M}\left(\array{c}
\chi_h\\0\endarray\right),
\label{Spinoren1}
\end{eqnarray}
with the nucleon Pauli spinors
$\chi_{\uparrow}=\left(\array{c}1\\0\endarray\right)$ and
 $\chi_{\downarrow}= \left(\array{c}0\\1\endarray\right)$.
After decomposing the vertex function
\begin{equation}
\Gamma_{\beta}=\left(\array{cc}\Gamma_{\beta}^A&\Gamma_{\beta}^B\\
\Gamma_{\beta}^C&\Gamma_{\beta}^D\endarray
\right), 
\end{equation}
we find for the leading non-relativistic term:
\begin{eqnarray}\label{tr}
  \varepsilon^{\mu*}_+ \,Tr [\dots] \, \varepsilon^{\nu}_+
\approx 
(2M)^2  \sum_{h,h'} 
tr  \left[ \Gamma_{\beta}^A\,\chi_{h'}\chi_{h'}^{\dag}\,
\Gamma_{\alpha}^{A\dag}
\,\chi_{h}\chi_{h}^{\dag}\,
{\cal A}^{\gamma^* p}_{+h}
\right],
\end{eqnarray}
where the trace is taken in the $2\times2$ spin-space. 
\newline
We also expand the denominator in Eq.(\ref{single amp2}) 
and use in leading non-relativistic order: 
\begin{eqnarray} \label{aprox}
\frac{1}{2 E_p(p^2-M^2)^2} 
\approx \frac{1}{32 M^3 \left(B+\frac{\vek{p}^2}{2M}\right)^2}.
\end{eqnarray}
Combined with Eq.(\ref{tr}) we obtain:
\begin{eqnarray}\label{kurz amp}
{\cal A}^{(1)}_{+H}&=&\frac{1}{8M}\int\frac{d^3 p}{(2\pi)^3}\,
\frac{{\cal E}^{*\alpha}_{H}{\cal E}^{\beta}_{H}}
{\left(B+\frac{\vek{p}^2}{2M}\right)^2}\,
\sum_{h,h'} 
tr \left[ \Gamma_{\beta}^A\,\chi_{h'}\chi_{h'}^{\dag}\, 
\Gamma_{\alpha}^{A\dag}
\,\chi_{h}\chi_{h}^{\dag}\,
{\cal A}^{\gamma^* p}_{+h}
\right] + [p\leftrightarrow n].
\end{eqnarray}
We now specify the non-relativistic deuteron wave-function with 
helicity $H$. 
For a spectator nucleon (here neutron) with helicity $h'$ 
we find:
\begin{eqnarray}\label{ident1}
\psi^{\dag}_{H,h'}(\vek{p})&=&\frac{\chi_{h'}^{\dag}
{\cal E}^{\alpha*}_{H} \,\Gamma^{A\dag}_{\alpha}}
{\sqrt{8M} (B+\frac{\vek{p}^2}{2M})},\\
\psi_{H,h'}(\vek{p})&=&
\frac{\Gamma^A_{\beta}\,{\cal E}^{\beta}_{H}\chi_{h'}}
{\sqrt{8M}(B+\frac{\vek{p}^2}{2M})}\label{ident2}.
\end{eqnarray}
In Eq.(\ref{kurz amp}) one can express 
the sum over the proton helicities using helicity projection operators:
\begin{eqnarray}\label{tensorprodukt}
\sum_h\chi_{h}\chi_{h}^{\dag}{\cal
A}^{\gamma^*p}_{+h}&=&\chi_{\uparrow}\chi_{\uparrow}^{\dag}{\cal
A}^{\gamma^*p}_{+\uparrow}+\chi_{\downarrow}\chi_{\downarrow}^{\dag}{\cal
A}^{\gamma^*p}_{+\downarrow}=\left(\array{cc}1&0\\0&0\endarray 
\right){\cal
A}^{\gamma^*p}_{+\uparrow}+\left(\array{cc}0&0\\0&1\endarray\right){\cal
A}^{\gamma^*p}_{+\downarrow} \nonumber\\
&=&P_{\uparrow}^p{\cal A}^{\gamma^*p}_{+\uparrow}+
P_{\downarrow}^p{\cal A}^{\gamma^*p}_{+\downarrow}.
\end{eqnarray}
This gives:
\begin{eqnarray}
{\cal A}^{(1)}_{+H}&=&\int\frac{d^3 p}{(2\pi)^3}\,
\sum_{h'} tr \left[\psi_{H,h'}(\vek{p})\psi^{\dag}_{H,h'}(\vek{p})\,
\left( P^p_{\uparrow} {\cal A}^{\gamma^*p}_{+\uparrow}+P^p_{\downarrow}
{\cal A}^{\gamma^*p}_{+\downarrow} \right)\right] +[p\leftrightarrow n].
\end{eqnarray}
Summing over the neutron helicities 
and introducing the coordinate-space wave functions (\ref{psi}), this
finally leads to:
\begin{eqnarray} \label{single amp4}
{\cal A}^{(1)}_{+H}&=&\int\frac{d^3 p}{(2\pi)^3}\;\int d^3r' \, 
e^{i\vek{p}\cdot\vek{r'}}\,\int d^3r \,e^{-i\vek{p}\cdot\vek{r}}
\psi^{\dag}_{H}(\vek{r'})
\left( P^p_{\uparrow}{\cal A}^{\gamma^*p}_{+\uparrow}+P^p_{\downarrow}
{\cal A}^{\gamma^*p}_{+\downarrow} \right)
\psi_{H}(\vek{r})  + [p\leftrightarrow n],
\nonumber \\
&=&\int d^3r\;
\psi^{\dag}_{H}(\vek{r})
\left( P^p_{\uparrow}{\cal A}^{\gamma^*p}_{+\uparrow} +P^p_{\downarrow}
{\cal A}^{\gamma^*p}_{+\downarrow} \right)
\psi_{H}(\vek{r})
+ \,[p\leftrightarrow n].
\end{eqnarray}
\subsection{Double scattering}
\label{App_B}

Here we start from the double scattering amplitude of Eq.(\ref{eq:A2_ds}): 
\begin{eqnarray}
i {\cal A}^{(2)}_{+H}&=&
-\sum_X\int\frac{d^4 p}{(2\pi)^4}
\int\frac{d^4k}{(2\pi)^4}
{\cal E}^{\alpha*}_{H}\varepsilon^{\mu*}_+
\frac{-i(g^{\rho\sigma}-p_X^{\rho}p_X^{\sigma}/M_X^2)}
{p_X^2-M_X^2+i\varepsilon}
\nonumber\\
&\times& Tr\left[\Gamma_{\beta}
(p_d,p)\,
\frac{i}{\Slash p_d-\Slash p -M+i\varepsilon}
i\,\hat{t}^{Xn\to \gamma^*n}_{\mu\rho}\,
\frac{i}{\Slash p_d-\Slash p- \Slash k-M+i\varepsilon}
\bar{\Gamma}_{\alpha}(p_d,p)\,
\right.\nonumber\\
&&\hspace*{0.5cm}
\times \left.\frac{i}{\Slash p +\Slash k-M+i\varepsilon}i\,
\hat{t}^{\gamma^*p\to Xp}_{\sigma\nu}\,
\frac{i}{\Slash p-M+i\varepsilon}\,
\right]
{\cal E}^{\beta}_{H}\varepsilon^{\nu}_+\nonumber\\
&+&[p\leftrightarrow n],
\end{eqnarray}
with the reduced diffractive production amplitudes $\hat{t}$ as 
specified in
Eq.(\ref{freie2}).
As in the single scattering case 
we  perform the energy integration assuming that all relevant poles are
in the nucleon propagators. Picking up the positive energy poles this is
equivalent to the replacements:
\begin{eqnarray} \label{mass shell2}
\frac{1}{(p_d-p)^2-M^2+i\varepsilon}\to-i\pi\frac{\delta(p_d^0-p^0-E_p)}
{E_p},\\
\frac{1}{(p+k)^2-M^2+i\varepsilon}\to-i\pi\frac{\delta(p^0+k^0-E_{p+k})}
{E_{p+k}},
\end{eqnarray}
with $E_{p+k}=\sqrt{M^2+(\vek{p}+\vek{k})^2}$.
We then obtain:
\begin{eqnarray}\label{double amp2}
{\cal A}^{(2)}_{+H}&=&-\sum_X
\int\frac{d^3p}{(2\pi)^3 2E_p}\int\frac{d^3k}{(2\pi)^3 2E_{p+k}}
\frac{(-g^{\rho\sigma}+p^{\rho}_Xp^{\sigma}_X/M_X^2)}
{p_X^2-M_X^2+i\varepsilon}\nonumber\\
&&\hspace*{1cm}\times\left.
\frac{{\cal E}^{\alpha*}_{H}\varepsilon^{\mu*}_+\,Tr[...]\,
{\cal E}^{\beta}_{H}\varepsilon^{\nu}_+}{(p^2-M^2+i\varepsilon)
((p_d-p-k)^2-M^2+i\varepsilon)}
\right|_{p_d^0-p^0=E_p;\;p^0+k^0=E_{p+k}}\nonumber\\
&&\hspace*{1cm} + \,[p\leftrightarrow n],
\end{eqnarray}
with: 
\begin{eqnarray}\label{Tr..}
Tr[...]&=&Tr\left[ \Gamma_{\beta} \,(\Slash p_d-\Slash p+M)\,
\hat{t}^{Xn\to \gamma^*n}_{\mu\rho}\,(\Slash p _d-\Slash p-\Slash k+M)\,
\bar{\Gamma}_{\alpha}\,(\Slash p+\Slash k+M)\,
\hat{t}^{\gamma^*p\to Xp}_{\sigma\nu}
\,(\Slash p+M)\,
\right].
\end{eqnarray}
In the numerator we neglect again terms of order
$\vek{p}^2/M^2$ and replace 
$\protect{\Slash p} + M \approx \sum_{h} u(p,h) \bar{u}(p,h)$ 
etc. This leads to:
\begin{eqnarray} \label{eq:Tr[...]_2}
Tr[...]&\approx &\sum_{h,h'} Tr\left[\Gamma_{\beta}
\,u(p_d-p,h')\,t^{Xn\to\gamma^*n}_{\mu\rho }(k,h')\,\bar{u}(p_d-p-k,h')\,
\right.\nonumber\\
&\times&\left.
\bar{\Gamma}_{\alpha}\,u(p+k,h)\,t^{\gamma^*p\to Xp}_{\sigma\nu}(k,h)\,
\bar{u}(p,h)
\right].
\end{eqnarray}
Here we have assumed s-channel helicity conservation for  
high-energy diffraction.
The corresponding amplitudes are:   
\begin{equation}
t^{\gamma^*N\to XN}_{\sigma\nu}(k,h)=\bar{u}(p+k,h)\,
\hat{t}^{\gamma^*N\to XN}_{\sigma\nu}\,u(p,h). 
\end{equation}
They are  related to diffractive (virtual) photoproduction 
amplitudes by:
\begin{equation}\label{difamp}
T^{NX}_{+h}(k)=\varepsilon^{\sigma *}_{X+} \,t^{\gamma^*N\to
XN}_{\sigma\nu}(k,h) \,\varepsilon^{\nu}_{+},
\end{equation}
with $\varepsilon_{X+}$, the transverse polarization vector of the produced
hadron. 
Expanding the Dirac spinors in Eq.(\ref{eq:Tr[...]_2}) 
following Eq.(\ref{Spinoren1}) we arrive at:
\begin{equation}
Tr[...] \approx 
(2M)^2 \sum_{h,h'} tr\left[  \Gamma_{\beta}^A  \;\chi_{h'} 
t^{Xn\to \gamma^*n}_{\mu\rho}(k,h')\;\chi^{\dag}_{h'}\;
\Gamma_{\alpha}^{A\dag}\;\chi_{h}\; t^{\gamma^*p\to
Xp}_{\sigma\nu}(k,h)\;\chi^{\dag}_{h}\right].
\end{equation}
When taking also the energy denominators in the leading order 
non-relativistic limit, i.e.
\begin{eqnarray}
\frac{1}{2E_p(p^2-M^2)}&\approx&\frac{1}{-8M^2(B+\frac{\vek{p}^2}{2M})},\\
\frac{1}{2E_{p+k}((p_d-p-k)^2-M^2)}&\approx&
\frac{1}{-8M^2(B+\frac{(\vek{p} + \vek{k})^2}{2M})},
\end{eqnarray}
we can again identify the non-relativistic deuteron wave-functions. 
For a deuteron with helicity $H$ and a re-scattering nucleon 
(here neutron) with  polarization $h'$ we have:
\begin{eqnarray}
\psi^{\dag}_{H,h'}(\vek{p}+\vek{k})&=&
\frac{\chi_{h'}^{\dag}{\cal E}^{\alpha*}_H \,\Gamma^{A\dag}_{\alpha}}
{\sqrt{8M}\left(B+\frac{(\vek{p}+\vek{k})^2}{2M}\right)},\\
\psi_{H,h'}(\vek{p})&=&\frac{
\Gamma^A_{\beta}\,{\cal E}^{\beta}_H \chi_{h'}}
{\sqrt{8M}\left(B+\frac{\vek{p}^2}{2M}\right)}.
\end{eqnarray}
In terms of coordinate-space wave functions we find:
\begin{eqnarray} \label{eq:A2}
{\cal A}^{(2)}_{+H}&=&
-\frac{1}{2M} \sum_X 
\int\frac{d^3p}{(2\pi)^3}\int\frac{d^3k}{(2\pi)^3}\int d^3r'
e^{i(\vek{p}+\vek{k})\cdot\vek{r'}}\;\int d^3r
e^{-i\vek{p}\cdot\vek{r}}\frac{(-g^{\rho\sigma}+
p^{\rho}_Xp^{\sigma}_X/M_X^2)}{p_X^2-M_X^2+i\varepsilon}\nonumber\\
&\times&
{\cal E}^{\alpha*}_{H}
\varepsilon^{\mu *}_+
\sum_{h,h'}
tr\left[\psi_{H,h'}(\vek{r}) 
t^{Xn\to \gamma^*n}_{\mu\rho}(k,h')
\,\psi^{\dag}_{H,h'}(\vek{r'})\, \chi_{h}\,
t^{\gamma^*p\to Xp}_{\sigma\nu}(k,h)\,\chi^{\dag}_{h}\,
\,\right]
{\cal E}^{\beta}_{H}
\varepsilon^{\nu }_+
\nonumber\\
&+& [p\leftrightarrow n].
\end{eqnarray}
Neglecting the dependence of the diffractive amplitudes 
$t^{\gamma^*N\to XN}$ on the longitudinal momentum transfer $k_z$,  
we perform the $k_z$-integration, using $p_X=q-k$ and 
$q^{\mu}=(q_0,{\bf 0}_{\perp},q_z = \sqrt{q_0^2 + Q^2})$:
\begin{eqnarray}\label{teta}
\int_{-\infty}^{\infty}dk_z\frac{e^{ik_z z}}
{p_X^2-M_X^2+i\varepsilon}&=&\int_{-\infty}^{\infty}dk_z
\frac{e^{ik_z z}}
{(q_0-k_0)^2-(q_z-k_z)^2-\vek{k}_{\perp}^2-M_X^2+i\varepsilon}, 
\nonumber\\
&\approx &\Theta(-z)\frac{-i\pi}{q_0}e^{i\frac{z}{\lambda}}+
\Theta(z)\frac{-i\pi}{q_0}e^{i(2q_z z- \frac{z}{\lambda})}.  
\end{eqnarray}
Here we neglect terms proportional to the 
energy transfer $k_0$  and to the transverse momentum transfer 
$\vek k_{\perp}$ which are suppressed by the deuteron wave function 
in (\ref{eq:A2}). 
Furthermore we  have introduced  the coherence length 
\begin{eqnarray}\label{lambda}
\lambda&\approx &\frac{2q_0}{M_X^2+Q^2}.
\end{eqnarray}
Now we  make use of the completeness relation for the 
diffractively produced states:
\begin{equation}
-g^{\rho\sigma}+p^{\rho}_Xp^{\sigma}_X/M_X^2=\sum_{\lambda}
\varepsilon^{\rho}_{X\lambda}\varepsilon^{\sigma *}_{X\lambda}, 
\end{equation}
and insert the diffractive amplitudes (\ref{difamp}):
\begin{eqnarray}
{\cal A}^{(2)}_{+H}&=&\frac{i}{4Mq_0} \sum_X
\int\frac{d^2k_{\perp}}{(2\pi)^2}\int d^2 b\,
e^{i\vek{k}_{\perp}\cdot\vek{b}}
\int_{-\infty}^0dz\,
e^{i\frac{z}{\lambda}}\nonumber\\
&\times&\sum_{h,h'}tr\left[\psi_{H,h'}(\vek{r})\, 
T^{nX}_{+ h'}(k)\,\psi^{\dag}_{H,h'}(\vek{r})\,\chi_{h}\, 
T^{pX}_{+ h}(k)\,\chi^{\dag}_{h}\,\right]\nonumber\\
&+& [p\leftrightarrow n].
\end{eqnarray}
We have omitted the second term in Eq.(\ref{teta}) which vanishes 
for large photon energies.
Expressed in terms of helicity projection operators
we find:
\begin{eqnarray}
{\cal A}^{(2)}_{+H}&=&\frac{i}{4Mq_0}\sum_X
\int\frac{d^2k_{\perp}}{(2\pi)^2}\int d^2 b\,
e^{i\vek{k}_{\perp}\cdot\vek{b}}
\int_{-\infty}^0dz\,
e^{i\frac{z}{\lambda}}\nonumber\\
&\times&\psi^{\dag}_{H}(\vek{r})
\left(
P^p_{\uparrow}T^{pX}_{+\uparrow}(k)+P^p_{\downarrow}
T^{pX}_{+\downarrow}(k)\right)\otimes\left(
P^n_{\uparrow}T^{nX}_{+\uparrow}(k)+P^n_{\downarrow}
T^{nX}_{+\downarrow}(k)\right)
\psi_{H}(\vek{r}) \nonumber\\
&+&[p\leftrightarrow n],
\end{eqnarray}
where $\otimes$ indicates a product of matrizes operating independently 
on the proton and the neutron. 
In terms of the deuteron wave function (\ref{psi}) 
one obtains for the helicity dependent amplitudes:
\begin{mathletters}
\label{eq:AppB_A2}
\begin{eqnarray}
{\cal A}_{+ +} &=& \frac{i}{4Mq_0}\sum_X
\int\frac{d^2k_{\perp}}{(2\pi)^2}\int d^2 b\,
e^{i\vek{k}_{\perp}\cdot\vek{b}}
\int_{-\infty}^0dz\,
e^{i\frac{z}{\lambda}}\;\frac{1}{\vek{r}^2}\nonumber\\
&\times&\left\{\left[u^2|Y_{00}|^2
+uv\frac{1}{\sqrt{10}}\left(Y_{00}^*Y_{20}+c.c.\right)
+v^2\frac{1}{10}|Y_{20}|^2\right]
T^{pX}_{+\uparrow}(k)T^{nX}_{+\uparrow}(k)
\right.\nonumber\\
&&
+v^2\frac{3}{20}|Y_{21}|^2
\left[T^{pX}_{+\uparrow}(k)T^{nX}_{+\downarrow}(k)
+T^{pX}_{+\downarrow}(k)T^{nX}_{+\uparrow}(k)\right]
\left.
+v^2\frac{3}{5}|Y_{22}|^2 
\,T^{pX}_{+\downarrow}(k)T^{nX}_{+\downarrow}(k)\right\},
\\
{\cal A}_{+ -} &=& \frac{i}{4Mq_0}\sum_X
\int\frac{d^2k_{\perp}}{(2\pi)^2}\int d^2 b\,
e^{i\vek{k}_{\perp}\cdot\vek{b}}
\int_{-\infty}^0dz\,
e^{i\frac{z}{\lambda}}\;\frac{1}{\vek{r}^2}\nonumber\\
&\times&\left\{\left[u^2|Y_{00}|^2
+uv\frac{1}{\sqrt{10}}\left(Y_{00}^*Y_{20}+c.c.\right)
+v^2\frac{1}{10}|Y_{20}|^2\right]
T^{pX}_{+\downarrow}(k)T^{nX}_{+\downarrow}(k)
\right.\nonumber\\
&&
+v^2\frac{3}{20}|Y_{21}|^2
\left[T^{pX}_{+\uparrow}(k)T^{nX}_{+\downarrow}(k)
+T^{pX}_{+\downarrow}(k)T^{nX}_{+\uparrow}(k)\right]
\left.+v^2\frac{3}{5}|Y_{22}|^2 
\,T^{pX}_{+\uparrow}(k)T^{nX}_{+\uparrow}(k)\right\},
\\
{\cal A}_{+ 0} &=& \frac{i}{4Mq_0}\sum_X
\int\frac{d^2k_{\perp}}{(2\pi)^2}\int d^2 b\,
e^{i\vek{k}_{\perp}\cdot\vek{b}}
\int_{-\infty}^0dz\,
e^{i\frac{z}{\lambda}}\;\frac{1}{2\vek{r}^2}\nonumber\\
&\times&\left\{\left[u^2|Y_{00}|^2
-uv\frac{2}{\sqrt{10}}\left(Y_{00}^*Y_{20}+c.c.\right)
+v^2\frac{2}{5}|Y_{20}|^2\right]
\left[T^{pX}_{+\uparrow}(k)T^{nX}_{+\downarrow}(k)
+T^{pX}_{+\downarrow}(k)T^{nX}_{+\uparrow}(k)\right]
\right.
\nonumber\\
&&\left.+v^2\frac{3}{5}|Y_{21}|^2
\left[T^{pX}_{+\uparrow}(k)T^{nX}_{+\uparrow}(k)
+ 
T^{pX}_{+\downarrow}(k)T^{nX}_{+\downarrow}(k)\right]
\right\}.
\end{eqnarray}
\end{mathletters}
On the other hand the form factors for the polarized deuteron 
\begin{equation} 
S_{H}(\vek k)=\int
d^3r\,|\psi_{H}(\vek r )|^2 e^{i\vek k \cdot \vek r} 
\end{equation}
read:
\begin{mathletters}
\label{eq:AppB_SH}
\begin{eqnarray}
S_{+}(\vek k) &=& S_{-}(\vek k) = 
\int
d^3r\,e^{i \vek k \cdot \vek r}\frac{1}{\vek{r}^2}\nonumber\\
&\times&
\left\{u^2|Y_{00}|^2+uv\frac{1}{\sqrt{10}}\left(Y_{00}^* Y_{20}+c.c.\right)
+v^2\left[\frac{1}{10}|Y_{20}|^2
+\frac{3}{5}|Y_{22}|^2+\frac{3}{10}|Y_{21}|^2\right]\right\},
\\
S_{0}(\vek k) &=& \int
d^3r\,e^{i\vek k \cdot \vek r}\frac{1}{\vek{r}^2}\nonumber\\
&\times&\left\{u^2
|Y_{00}|^2 -uv\frac{2}{\sqrt{10}}\left(Y_{00}^*
Y_{20}+c.c.\right) +v^2\frac{2}{5}|Y_{20}|^2
+\frac{3}{5}v^2 |Y_{21}|^2\right\}.
\end{eqnarray}
\end{mathletters}
%

%
%
\begin{figure}
\vspace*{3cm}
\centering{\ \epsfig{figure=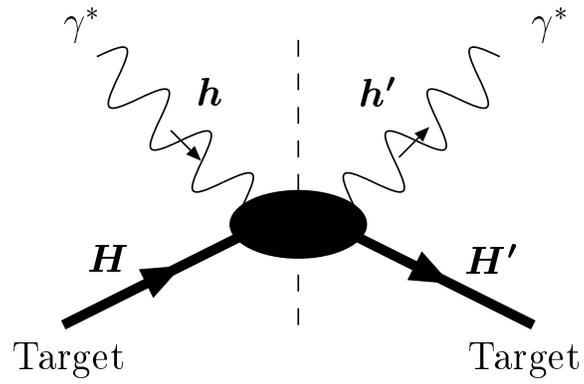,height=5cm}}
\vspace*{1cm}
\caption{
Forward helicity amplitude for Compton scatttering on a polarized target.}
\end{figure}
\pagebreak

%
\begin{figure}
\vspace*{3cm}
\centering{\ \epsfig{figure=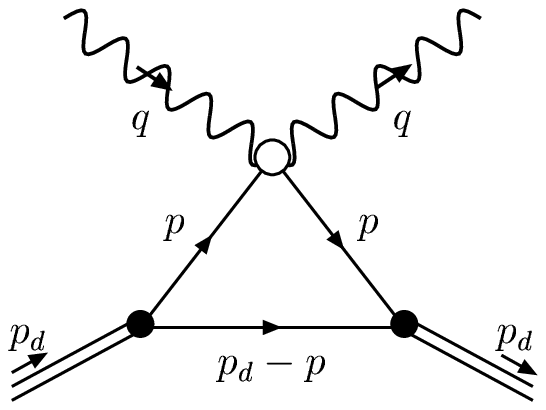,height=5cm}}
\vspace*{1cm}
\begin{center}
{FIG. 2a.}
Single scattering contribution to virtual photon-deuteron scattering.
\end{center}


\vspace*{3cm}
\centering{\ \epsfig{figure=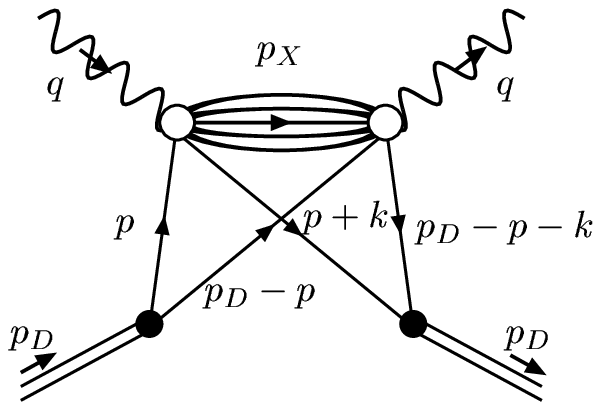,height=5cm}}
\vspace*{1cm}
\begin{center}
{FIG. 2b.}
Double  scattering contribution to virtual photon-deuteron scattering.
\end{center}
\end{figure}
\pagebreak


\setcounter{figure}{2}

\begin{figure}
\centerline{\epsfxsize=0.75\hsize \epsffile{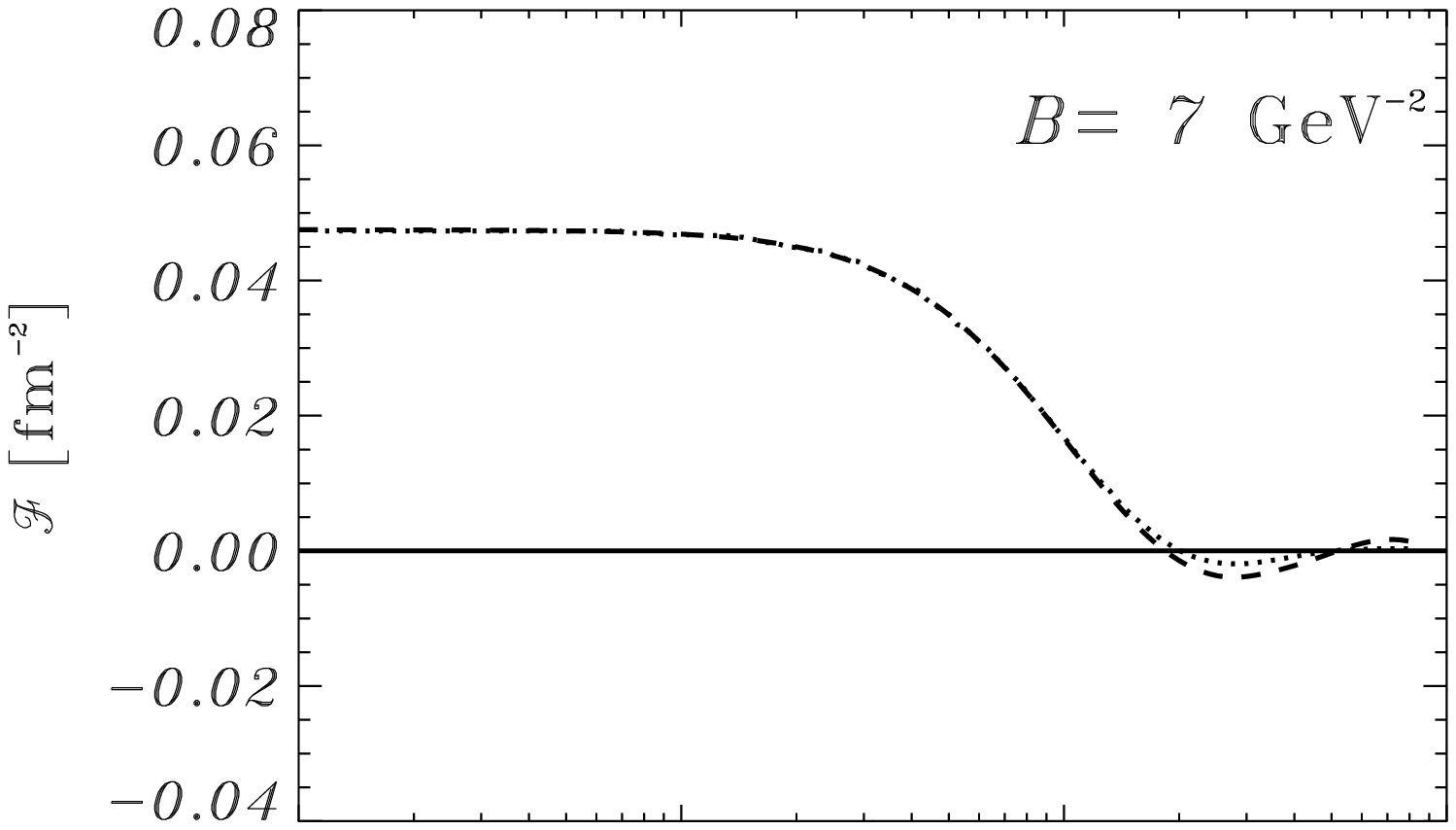}}
\vspace*{-2.5cm}
\centerline{\epsfxsize=0.75\hsize \epsffile{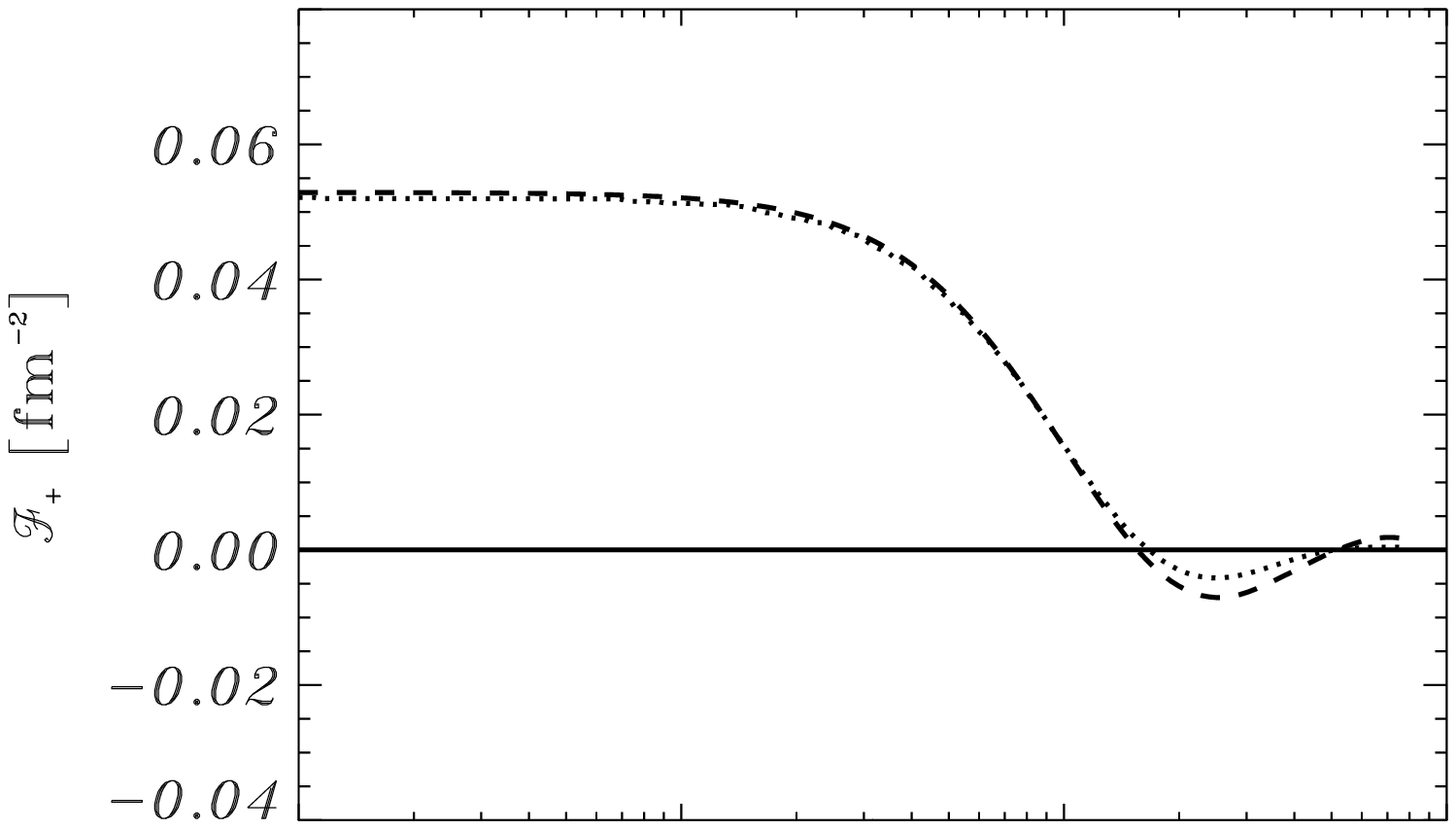}}
\vspace*{-2.5cm}
\centerline{\epsfxsize=0.75\hsize \epsffile{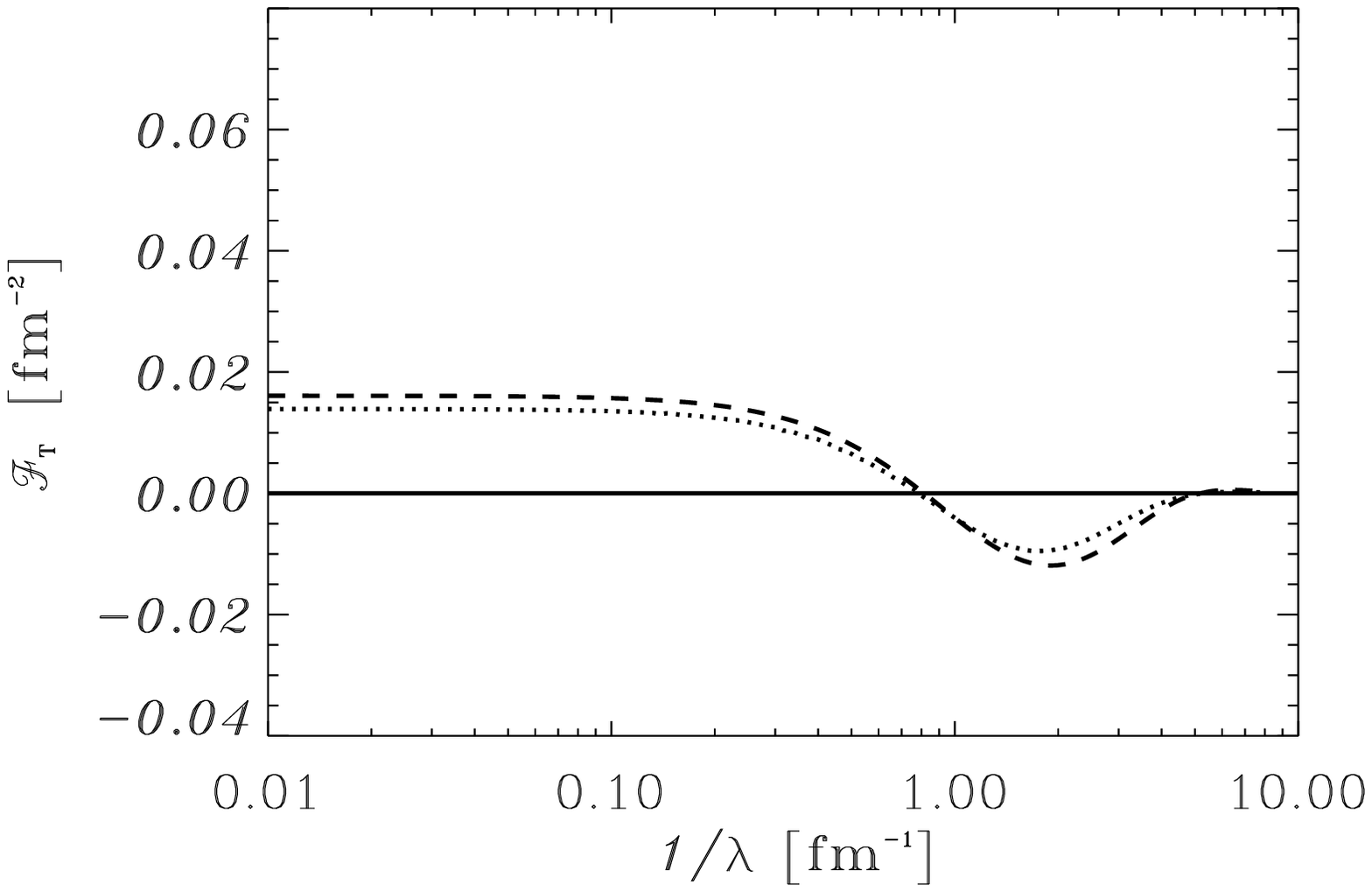}}
{FIG. 3a.}
Integrated deuteron form factor ${\cal F}_H$ from 
Eqs.(\ref{eq:F_H},\ref{eq:F_unpol},\ref{eq:F_02}) 
for different deuteron polarizations for an average 
slope $B = 7\,GeV^{-2}$.  
The dotted and dashed curves 
correspond to the Bonn [28] and Paris potential 
[27], respectively.
\end{figure}
\pagebreak

\begin{figure}
\centerline{\epsfxsize=0.75\hsize \epsffile{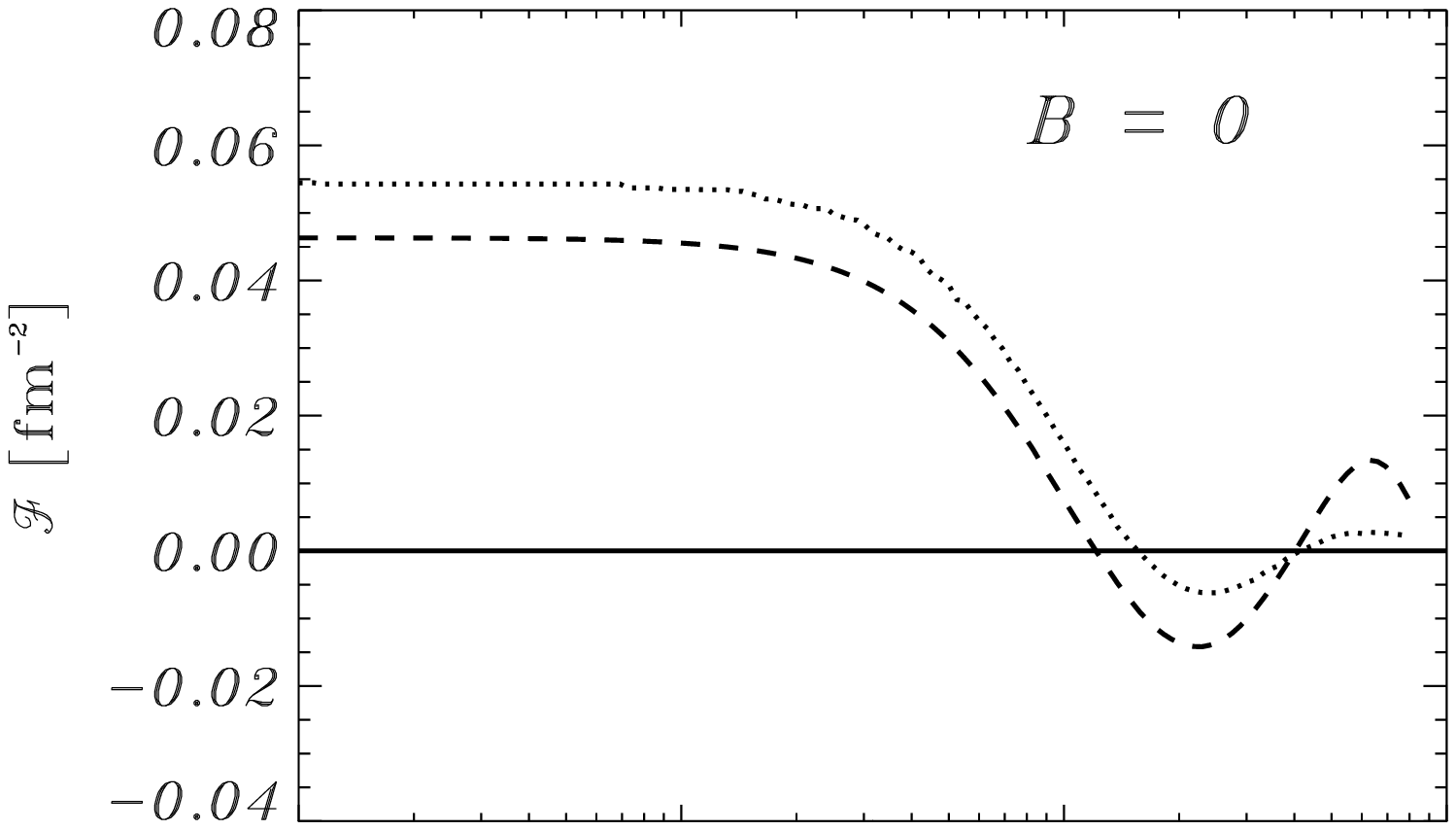}}
\vspace*{-2.5cm}
\centerline{\epsfxsize=0.75\hsize \epsffile{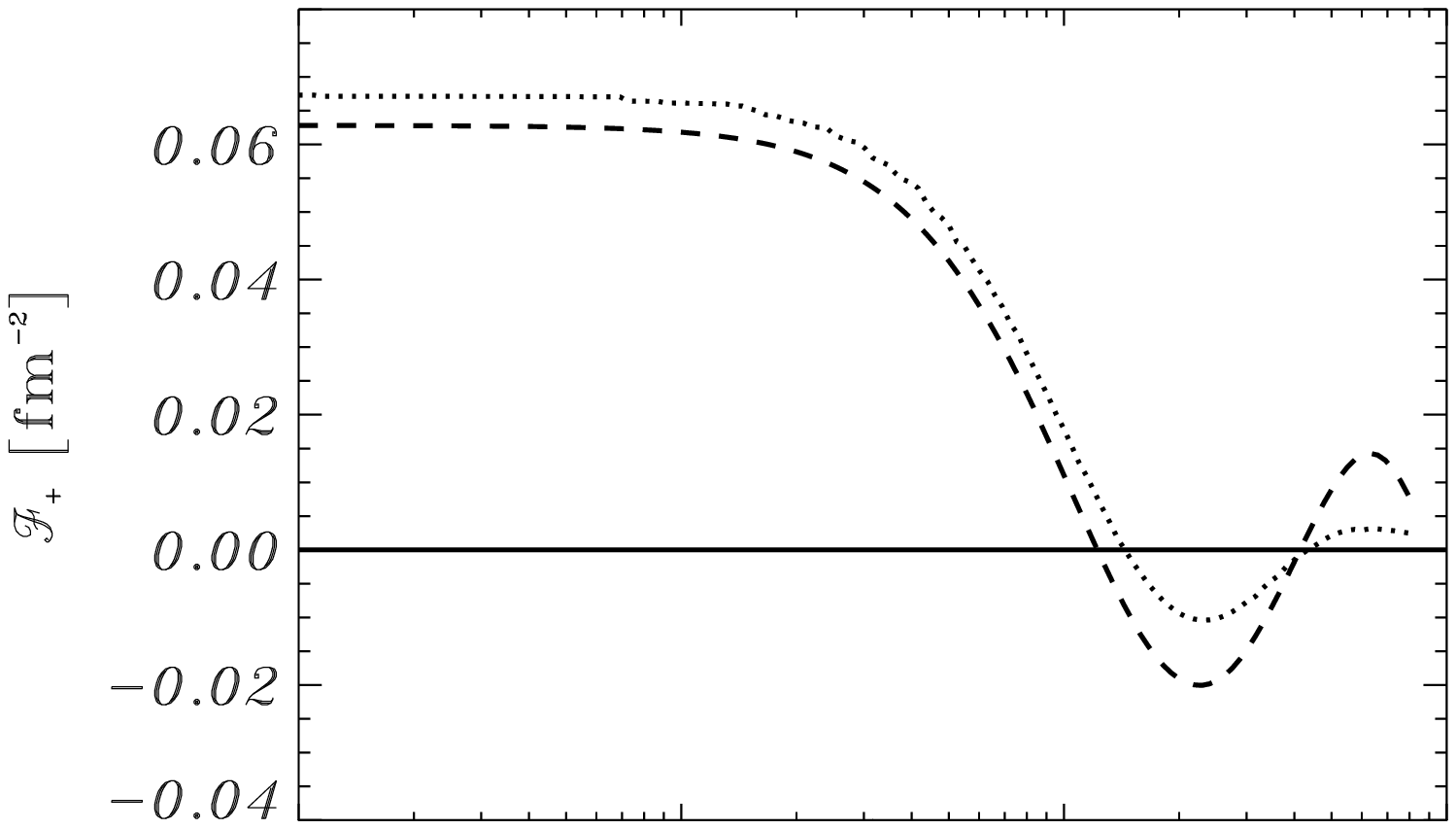}}
\vspace*{-2.5cm}
\centerline{\epsfxsize=0.75\hsize \epsffile{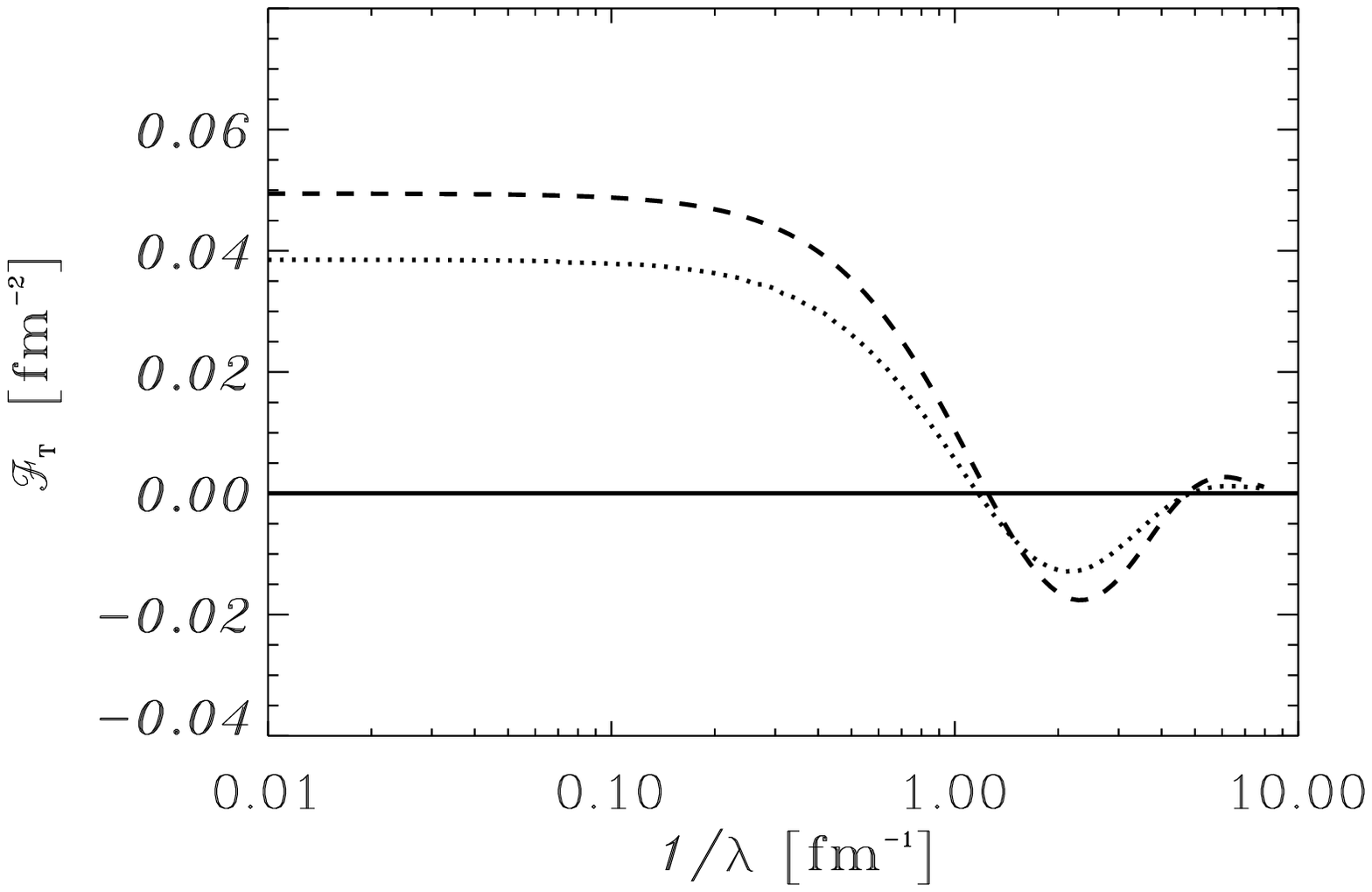}}
{FIG. 3b.}
Integrated deuteron form factor ${\cal F}_H$ from 
Eqs.(\ref{eq:F_H},\ref{eq:F_unpol},\ref{eq:F_02}) 
for different deuteron polarizations 
for an average slope $B = 0$.  
The dotted and dashed curves 
correspond to the Bonn [28] and Paris potential 
[27], respectively.
\end{figure}
\pagebreak

\setcounter{figure}{3}

\begin{figure}
\centering{\ \epsfig{figure=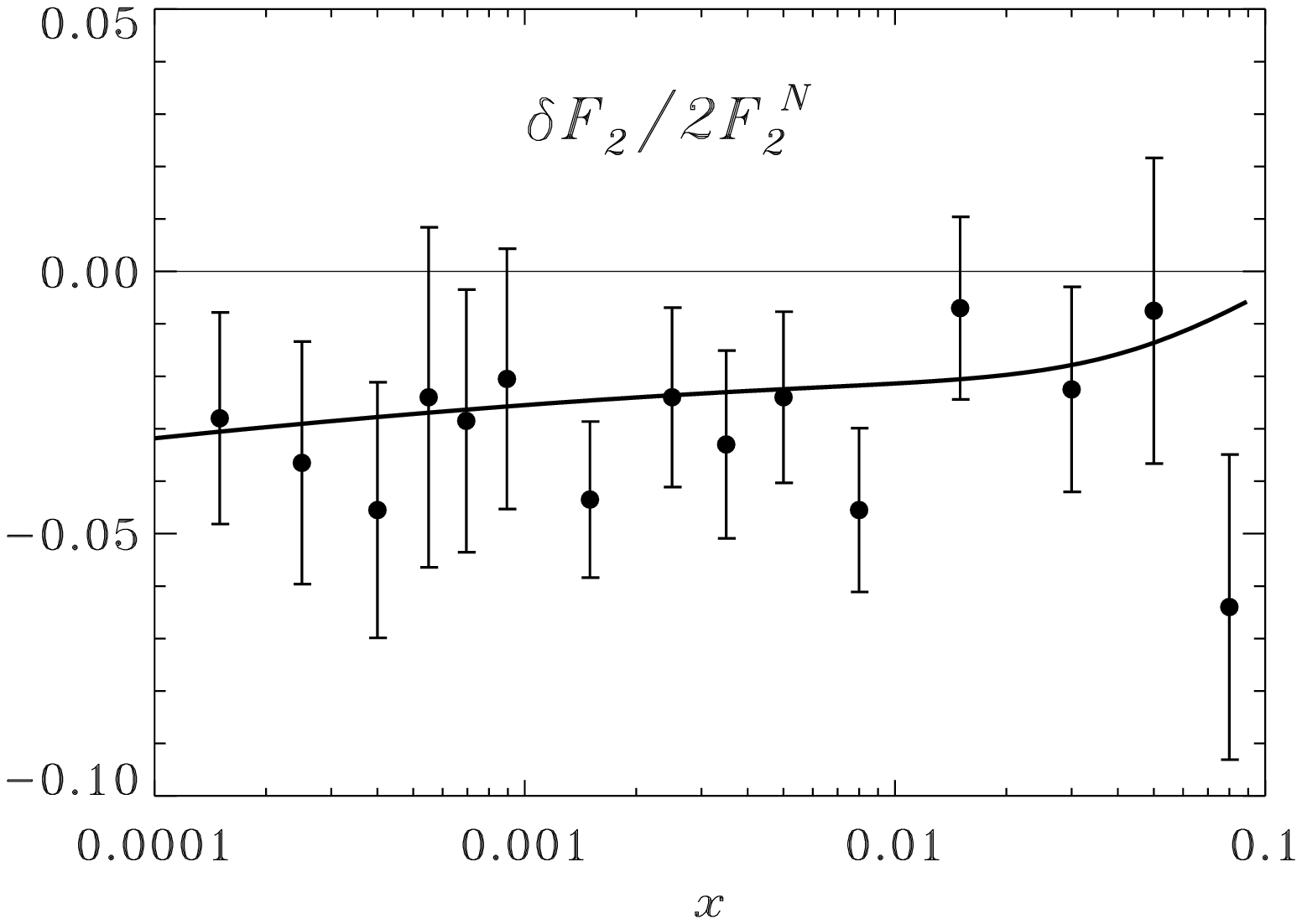,height=10cm}}
\caption
{Shadowing correction $\delta F_2/2F_2^N$ for deuterium, with data 
from E665 [6]. 
The full line
represents a fit to the data used in (43) and 
(44).}
\end{figure}
\pagebreak


\begin{figure}
\centering{\ \epsfig{figure=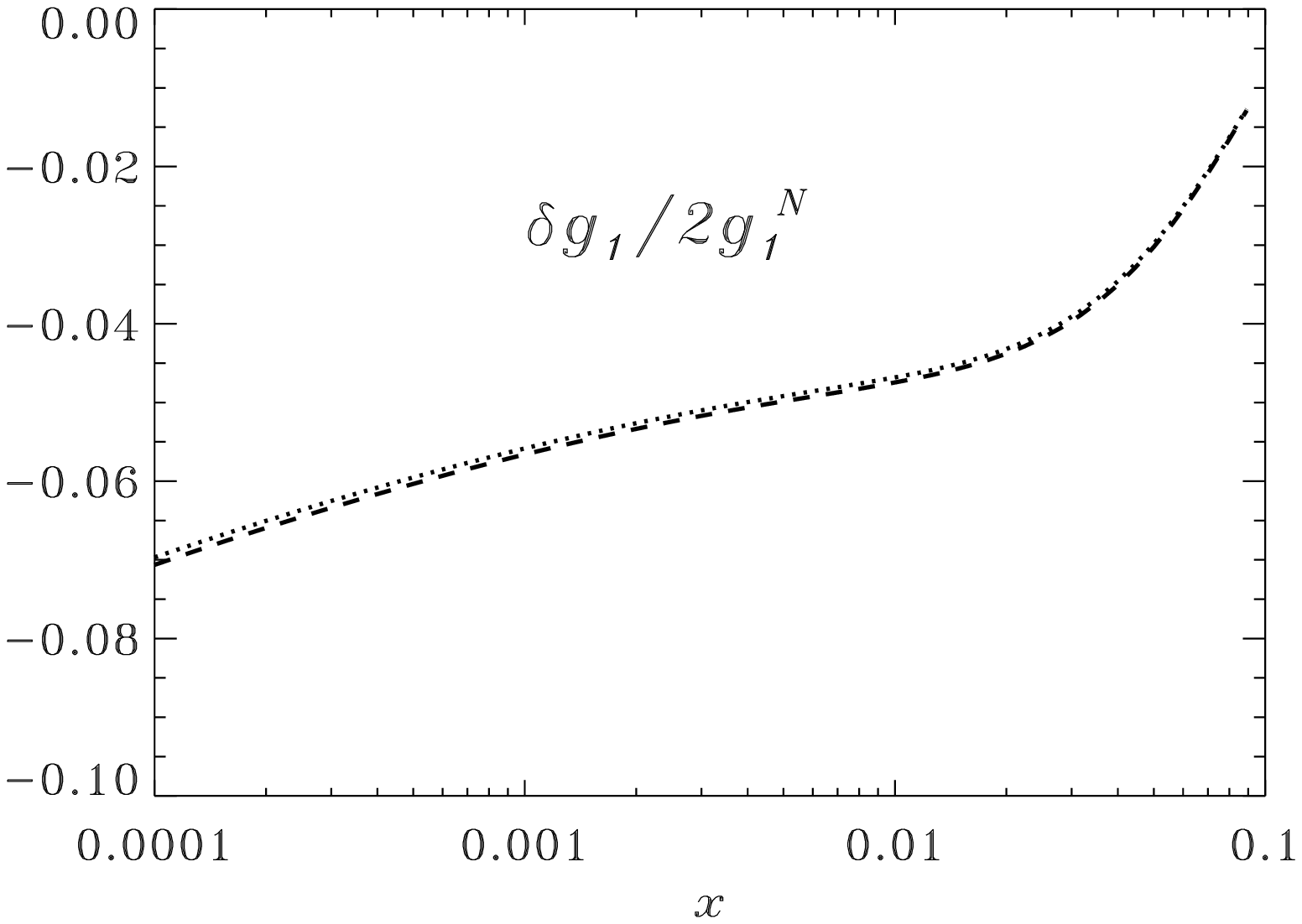,height=10cm}}
\caption
{Shadowing correction $\delta g_1/2g_1^N$ for deuterium. The dotted and 
dashed curves
correspond to the Bonn OBE [28] and Paris potential [27],  
respectively.}
\end{figure}
\pagebreak


\begin{figure}
\centering{\ \epsfig{figure=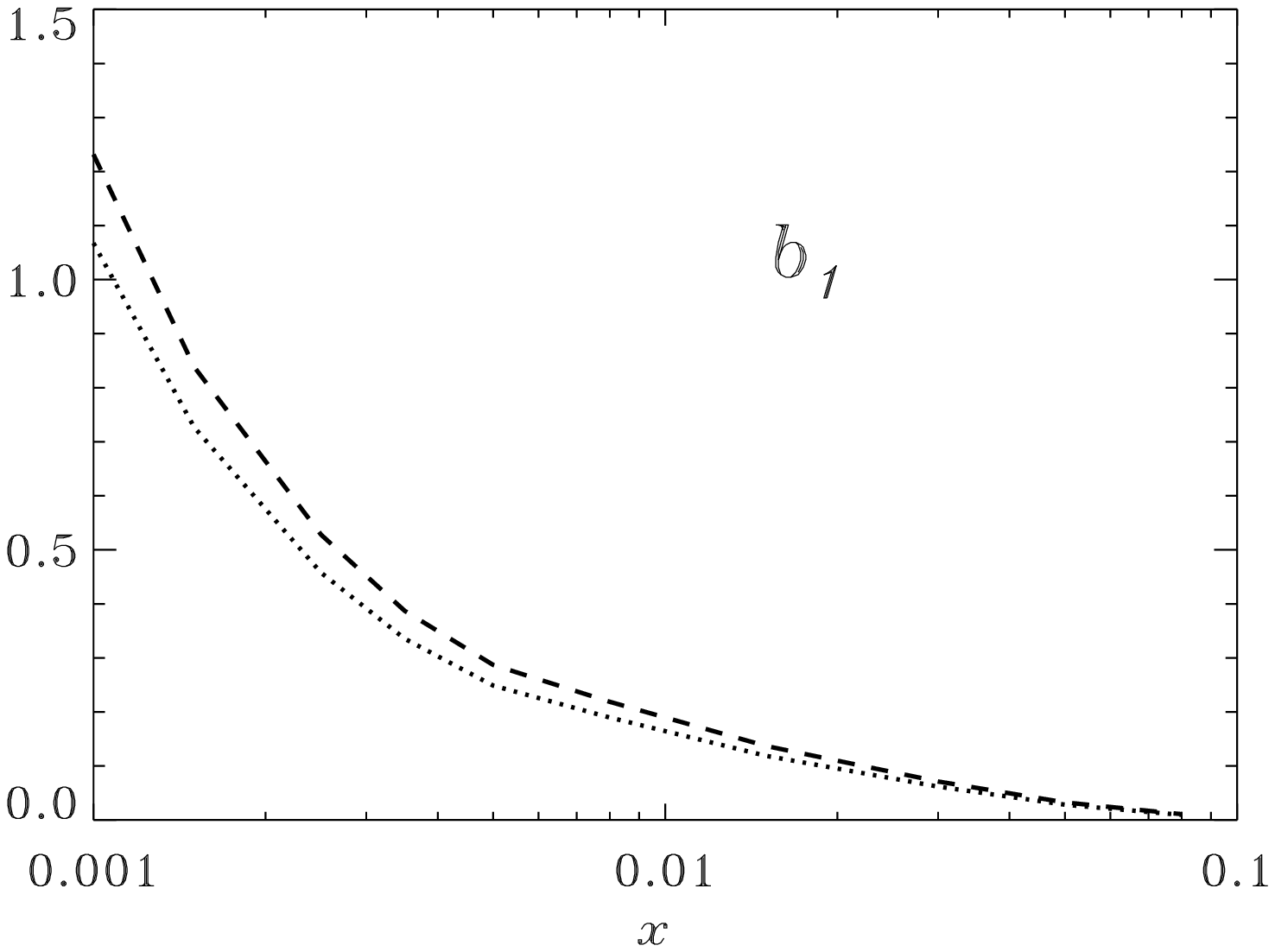,height=10cm}}
\caption
{Double scattering contribution to the tensor structure function 
$b_1$ of the deuteron. The dotted and dashed curves
correspond to the Bonn OBE [28] and Paris potential [27], 
respectively.}
\end{figure}
\pagebreak

\newpage














\end{document}